\documentclass[reprint, showpacs, groupedaddress
 amsmath,amssymb,
 aps,
 natbib,
]{revtex4-1}

\usepackage{graphicx}
\usepackage{dcolumn}
\usepackage{bm}
\usepackage{hyperref}
\usepackage{amssymb}
\usepackage{mathrsfs}
\usepackage{amsmath,fleqn}

\newcommand{\bea}{\begin{eqnarray}}
\newcommand{\eea}{\end{eqnarray}}
\newcommand{\ba}{\begin{array}}
\newcommand{\ea}{\end{array}}

\newcommand{\be}{\begin{equation}}

\newcommand{\ee}{\end{equation}}
\newcommand{\bt}{\begin{teo}}
\newcommand{\et}{\end{teo}}

\newcommand{\om}{\omega}

\newcommand{\bro}{\beta_{\rm BR}}
\newcommand{\blo}{\beta_{\rm loc}}

\newcommand{\biz}{\beta_{\rm IZ}}
\newcommand{\La}{\Lambda}
\newcommand{\linf}{{\cal L}}

\begin{document}

\preprint{APS/123-QED}

\title{Studies of dynamical localization in a finite-dimensional model of the quantum kicked rotator }

\author{Thanos Manos}
\email{thanos.manos@uni-mb.si}
\affiliation{CAMTP - Center for Applied Mathematics and Theoretical Physics, University of Maribor, Krekova 2, SI-2000 Maribor, Slovenia}
\affiliation{School of Applied Sciences, University of Nova Gorica, Vipavska 11c, SI-5270 Ajdov\v s\v cina, Slovenia}
\affiliation{Institute of Neuroscience and Medicine Neuromodulation
(INM-7), Research Center J\"ulich, D-52425 J\"ulich, Germany}

\author{Marko Robnik}
\email{Robnik@uni-mb.si}
\affiliation{%
 CAMTP - Center for Applied Mathematics and Theoretical Physics, University of Maribor, Krekova 2, SI-2000 Maribor, Slovenia}

\date{\today}

\begin{abstract}
We review our recent works on the dynamical localization in the quantum kicked rotator (QKR) and the related properties of the classical kicked rotator  (the standard map, SM).  We introduce the Izrailev $N$-dimensional model of the QKR and analyze the localization properties of the Floquet eigenstates [{\em Phys. Rev. E} {\bf 87}, 062905 (2013)], and the statistical properties of the quasienergy spectra. We survey normal and anomalous diffusion in the SM, and the related accelerator modes [{\em Phys. Rev. E} {\bf 89}, 022905 (2014)]. We analyze the statistical properties [{\em Phys. Rev. E} {\bf 91},042904 (2015)] of the localization measure, and show that the reciprocal localization length has an almost Gaussian distribution which has a finite variance even in the limit of the infinitely dimensional model of the QKR, $N\rightarrow \infty$. This sheds new light on the relation between the QKR and the Anderson localization phenomenon in the one-dimensional tight-binding model. It explains the so far mysterious strong fluctuations in the scaling properties of the QKR. The reason is that the finite bandwidth approximation of the underlying Hamilton dynamical system in the Shepelyansky picture [{\em Phys. Rev. Lett.} {\bf 56}, 677 (1986)] does not apply rigorously. These results call for a more refined theory of the localization length in the QKR and in similar Floquet systems, where we must predict not only the mean value of the inverse of the localization length  but also its (Gaussian) distribution. We also numerically analyze the related behavior of finite time Lyapunov exponents in the SM and of the $2\times2$ transfer matrix formalism.


\end{abstract}

\pacs{05.45.Mt,05.45.Ac,05.60.Cd}
\keywords{Suggested keywords}
\maketitle


\section{Introduction} \label{intro}

Quantum chaos, or wave chaos, is the study of the phenomena in the quantum
domain which correspond to the classical chaos in the Hamiltonian systems
\cite{Stoe,Haake}. Although quantum motion (time evolution of the
wavefunctions) of bound systems with purely discrete energy spectrum is
ultimately (after a sufficiently long time, asymptotically) stable and
regular, in fact almost periodic, it exhibits many features of the
classical motion such as e.g. diffusion in a chaotic domain, for time
up to the Heisenberg time. The Heisenberg time, also called break time, is
an important time scale in any quantum system, and is given by $t_H=2\pi\hbar/\Delta E$, where $h=2\pi\hbar$ is the Planck constant and $\Delta E$ is the mean energy level spacing, such that the mean energy level density is $\rho(E) =1/\Delta E$. For time shorter than approximately  $t_H$ the quantum diffusion follows the classical chaotic diffusion, but is stopped at larger times, just due to the interference phenomena, which occur due to the wave nature of the underlying system, and are typically destructive. Pictorially speaking, for time up to $t_H$ the quantum system behaves as if its evolution operator has a continuous spectrum, like the classical one has in the chaotic regime, but at later times it senses the discreteness of the spectrum. If the quantum diffusion stops, while the classical chaotic diffusion continues, we speak about the {\em dynamical localization}, or {\em quantum localization} or {\em Chirikov localization}, first observed in time-dependent systems \cite{CCFI79} (see also e.g. \cite{Stoe,Haake}). Through the Fourier
transform connection between the time and energy, the dynamical localization
reflects itself also in the time-independent eigenfunctions, both in the eigenstates of the Floquet operator in time-periodic systems and in the eigenfunctions of the time-independent, classically chaotic, systems. Namely, if all classical transport times like the diffusion time (time necessary to occupy the entire classically available chaotic part of the phase space)
are all shorter than the Heisenberg time $t_H$, we find extended eigenstates, and localized eigenstates otherwise. The subject of this review is to summarize our main recent results on dynamical localization in time-dependent periodic (Floquet) systems, exemplified by the quantum kicked rotator (QKR), but the approach is nevertheless quite general.

Another important aspect of quantum chaos is the statistics of the energy spectra of classically chaotic quantal systems. One of the main cornerstones in the development of quantum chaos \cite{Stoe,Haake,Rob1998} is the finding that in classically fully chaotic, ergodic, autonomous Hamilton systems with the purely discrete spectrum the fluctuations of the energy spectrum around its mean behavior obey the statistical laws described by the Gaussian Random Matrix Theory (RMT) \cite{Mehta,GMW}, provided that we are in the sufficiently deep semiclassical limit. The latter condition means, as explained above,  that all relevant classical transport times are smaller than Heisenberg time $t_H$. This statement is known as the Bohigas -Giannoni - Schmit (BGS) conjecture and goes back to their pioneering paper in 1984 \cite{BGS}, although some preliminary ideas were published in \cite{Cas}. Since $\Delta E \propto \hbar^f$, where $f$ is the number of degrees of freedom (= the dimension of the configuration space), we see that for sufficiently small $\hbar$ the stated condition will always be satisfied. Alternatively, fixing the $\hbar$, we can go to high energies such that the classical transport times become smaller than $t_H$. The role of the antiunitary symmetries that classify the statistics in terms of GOE, GUE or GSE (ensembles of RMT) has been elucidated in \cite{RB1986}, see also \cite{Rob1986}, and \cite{Stoe,Haake,Rob1998,Mehta}. The theoretical foundation for the BGS conjecture has been initiated first by Berry \cite{Berry1985}, and later further developed by Richter and Sieber \cite{Sieber}, arriving finally in the almost-final proof proposed by the group of F. Haake \cite{Mueller1,Mueller2,Mueller3,Mueller4}.

On the other hand, if the system is classically integrable, Poisson statistics applies, as is well known and goes back to the work by Berry and Tabor in 1977 (see \cite{Stoe,Haake,Rob1998} and the references therein, and for the recent advances \cite{RobVeb}).

In the mixed type regime, where classical regular regions coexist in the classical phase space with the chaotic regions, being a typical KAM-scenario which is the generic situation, the so-called Principle of Uniform Semiclassical Condensation (of the Wigner functions of the eigenstates; PUSC) applies, based on the ideas by Berry \cite{Berry1977}, and further extended by
Robnik \cite{Rob1998}. Consequently the Berry-Robnik statistics \cite{BR1984,
ProRob1999}  is observed, again under the same  semiclassical condition stated above requiring that $t_H$ is larger than all classical transport times.

The relevant papers dealing with the mixed type regime after the work \cite{BR1984} are \cite{ProRob1993a,ProRob1993b,ProRob1994a,ProRob1994b,Pro1998a,Pro1998,GroRob1,GroRob2} and the most recent advance was published in \cite{BatRob2010,BMR2013,BatRob2013,BatRob2013A}. If the couplings between the regular eigenstates and chaotic eigenstates become important, due to the dynamical tunneling, we can use the ensembles of random matrices that capture these effects \cite{VSRKHG}.  As the tunneling strengths typically decrease exponentially with the inverse effective Planck constant, they rapidly disappear with increasing energy, or by decreasing the value of the Planck constant. In such case the regular and chaotic eigenstates can be separated and the dynamical localization in the chaotic eigenstates can be studied. For an excellent  review of dynamical localization in the time-independent systems see the paper by Prosen \cite{Pro2000} and the references therein. In such a situation it turns out that the Wigner functions of the chaotic eigenstates no longer uniformly occupy the entire classically accessible chaotic region in the classical phase space, but are localized on a proper subset of it.  Indeed, his has been analyzed with unprecedented precision and statistical significance by Batisti\'c and Robnik \cite{BatRob2010,BatRob2013,BatRob2013A} in case of mixed type systems. The important discovery is that the level spacing distribution of the dynamically localized chaotic eigenstates in periodic as well as time-independent systems is exceedingly well described by the Brody distribution, introduced in \cite{Bro1973}, see also \cite{Bro1981}, with the Brody parameter values $\bro$ within the interval $[0,1]$, where $\bro=0$ yields the Poisson distribution in case of the strongest localization, and $\bro=1$ gives the Wigner surmise (2D GOE, as an excellent approximation of the infinite dimensional GOE). The Brody distribution  was found to fit  the empirical data much better than the distribution function proposed by Izrailev (see \cite{Izr1988,Izr1990} and the references therein) characterized by the parameter $\biz$. It is well known that Brody distribution so far has no theoretical foundation, but our empirical results show that we have to consider it seriously thereby being motivated for seeking its physical foundation.

In this review of our papers \cite{MR2013,MR2014,MR2015} we explore the quantum kicked rotator (QKR) introduced by Casati \textit{et al.} \cite{CCFI79} from the classical point of view (the standard map, SM), analyze
the quantum analog using the $N$-dimensional model of Izrailev, and consider the semiclassical connection between the two pictures. We shall treat the cases with the classical dimensionless kick parameter $K$ in the range $K\in[5,35]$, and for some purposes even up to $K=70$, and in the end shall focus on
the case $K=10$, which is the most chaotic one in the sense that it is
fully chaotic with minimal (in fact undetected) regular regions among
all cases $K\in[5,70]$ and among them best exhibits normal diffusion.
Izrailev's $N$-dimensional model introduced and discussed in
\cite{Izr1986,Izr1987,Izr1989,Izr1990} is treated for various $N \le 3000$, which in the limit $N \rightarrow \infty$  tends to the QKR.  Due to the finiteness of $N$ the observed (dimensionless) localization length of the eigenfunctions in the space of the angular momentum quantum number does not possess a sharply defined value, but has a certain distribution instead.  Its reciprocal value is almost Gaussian distributed. This might be expected on the analogy with the finite time Lyapunov exponents in the Hamiltonian dynamical systems. In order to corroborate the theoretical findings on this topics we perform  the numerical analysis of the finite time Lyapunov exponents in the standard map (classical kicked rotator), especially the decay of the variance. Indeed, in the Shepelyansky picture \cite{She1986} the localization length can be obtained as the inverse of the smallest positive Lyapunov exponent of a finite $2k$-dimensional Hamilton system associated with the band matrix representation of the QKR, where $k$ is the quantum kick parameter (to be precisely defined below).  In this picture, $N$ plays the role of time. However, unlike the chaotic classical maps or products of transfer matrices in the Anderson tight-binding approximation, where the mean value of the finite time Lyapunov exponents is usually equal to their asymptotical value of infinite time and the variance decreases inversely with time, as we also carefully checked, here the distribution is found to be independent of $N$: It has a nonzero variance even in the limit $N\rightarrow \infty$. The reason is that the quantum kicked rotator at $N=\infty$ cannot be {\em exactly} modeled with finite bandwidth (equal to $2k$) band matrices, but only approximately, such that the underlying Hamilton system of the Shepelyansky picture has a growing dimension with $N$, implying asymptotically an infinite set of Lyapunov exponents and behavior different from the finite dimensional Hamiltonian systems. The observation of the distribution of the localization length around its mean value with finite variance also explains the strong fluctuations in the scaling laws of the kicked rotator, such as e.g. the entropy localization measure as a function of the theoretical scaling parameter $\Lambda$, to be discussed below. On the other hand, the two different empirical localization measures, namely the mean localization length as extracted directly from the exponentially localized eigenfunctions and the measure based on the information entropy of the eigenstates, are perfectly well linearly connected and thus equivalent. Therefore these results call for a refined theory of the localization length in the quantum kicked rotator and similar systems, where we must predict not only the mean value of the inverse localization length but also its (Gaussian) distribution, in particular the variance \cite{MR2015} (Manos and Robnik 2015).

The paper is organized as follows. In Sec.~\ref{sec1} we introduce and study the the classical kicked rotator (standard map, SM) and concentrate on the role of accelerator modes for the anomalous diffusion. In Sec.~\ref{sec2} we
introduce the quantum kicked rotator (QKR) and the $N$-dimensional Izrailev model and study the localization properties, including the scaling laws. In Sec.~\ref{sec3} we analyze the numerical results showing that the localization
measure has a distribution whose variance does not go to zero in the limit
$N \rightarrow \infty$. In Sec.~\ref{sec4} we discuss the results and draw
the main conclusions, pointing out the important differences between the QKR
and the one-dimensional Anderson localization in the tight-binding approximation.

\section{The  classical kicked rotator:  The standard map}
\label{sec1}

The kicked rotator was introduced by Casati \textit{et al.} \cite{CCFI79} and is one of the key model systems in classical and quantum chaos, especially for time-periodic (Floquet) systems.  The Hamiltonian function is
\be  \label{KR}
H= \frac{p^2}{2I} + V_0 \,\delta_T(t)\,\cos \theta.
\ee
Here $p$ is the (angular) momentum, $I$ the moment of inertia, $V_0$ is the strength of the periodic kicking, $\theta$ is the (canonically conjugate, rotation) angle, and $\delta_T(t)$ is the periodic Dirac delta function with period $T$. Between the kicks the rotation is free,  therefore the Hamilton equations of motion can be immediately integrated,  and thus the dynamics can be reduced to the standard map (SM), or so-called Chirikov-Taylor map, given by
\be \label{SM1}
 \left\{
 \begin{aligned} \label{SM1}
p_{n+1} &= p_n + V_0 \sin \theta_{n+1},\\
\theta_{n+1} &= \theta_n + \frac{T}{I} p_n,
 \end{aligned}
 \right.
\ee
and introduced in \cite{T69,F72,C79}. Here the quantities $(\theta_n, p_n)$ refer to their values just immediately after the $n$-th kick. Then,
by introducing new dimensionless momentum $P_n = p_nT/I$, we get
\be
 \left\{
 \begin{aligned} \label{SM2}
P_{n+1} &= P_n + K \sin \theta_{n+1},\\
\theta_{n+1} &= \theta_n  + P_n,
 \end{aligned}
 \right.
\ee
where the system is now governed by a single classical {\em dimensionless}
kick parameter $K=V_0 T/I$, and the mapping is area preserving.

The generalized diffusion process of the standard map [Eq.~(\ref{SM2})] is defined by
\be \label{varp}
\langle(\Delta P)^2\rangle = D_{\mu}(K) n^{\mu},
\ee
where $n$ is the number of iterations (kicks), and the exponent $\mu$ is in the interval $[0,2)$, and all variables $P$, $\theta$ and $K$ are dimensionless. Here $D_{\mu}(K)$ is the generalized classical diffusion constant. In the case $\mu=1$ we have the \textit{normal diffusion}, and $D_1(K)$ is then the normal diffusion constant, whilst in the case of anomalous diffusion we observe \textit{subdiffusion} when $0 < \mu < 1$  or \textit{superdiffusion} if $1 <\mu \le2$. In the case $\mu=2$ we have the \textit{ballistic transport} which is associated strictly with the presence of accelerator modes.

In the case of the normal diffusion $\mu=1$ the theoretical value of $D_1(K)$ is given in the literature, e.g. in \cite{Izr1990} or \cite{LL1992},
\begin{flalign} \label{Dcl}
 D_{1}(K)=
\begin{cases}
 \frac{1}{2} K^2\left [1- 2J_2(K) \left (1-J_2(K) \right ) \right ], \text{if} \ K \ge 4.5 \\
 0.15(K-K_{cr})^3, \text{if} \ K_{cr} < K \le 4.5
\end{cases},
\end{flalign}
where $K_{cr}\simeq 0.9716$ and $J_2(K)$ is the Bessel function. Here we neglect higher terms of order $K^{-2}$. However, there are many important subtle details in the classical diffusion further discussed below.

\begin{figure}\centering
\includegraphics[width=\columnwidth]{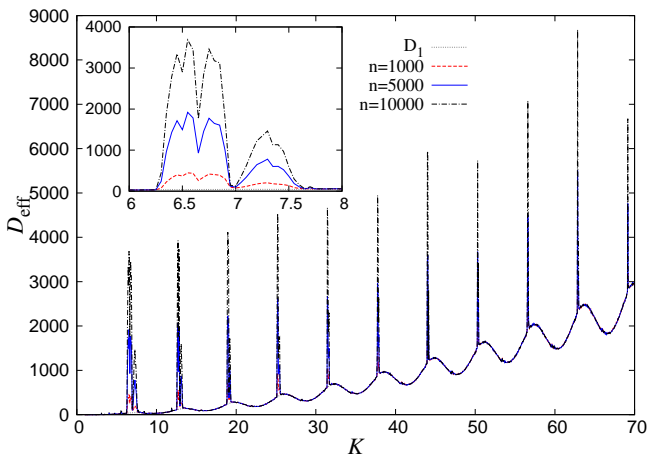}
\caption{(Color online) The classical diffusion constant  $D_{\rm eff}=\langle(\Delta P)^2\rangle/n$ for the standard map as a function of $K$ ($\delta K=0.05$) for three discrete times $n$, i.e., the number of the iterations of the standard map, $n=1000$ (lower red dashed line), $n=5000$ (intermediate blue solid) and $n=10000$ (upper black dot-dashed). In the background we have plotted the classical diffusion constant $D_1$ (gray dotted line) [Eq.~(\ref{Dcl})]. The presence of  accelerator modes at certain intervals of $K$ (and the sticky objects around) generate anomalous diffusion which is rendered by peaks. Here we used $\approx$100000 $(314 \times 314$) initial conditions uniformly distributed in a grid on the entire phase space $[0,2\pi]\times[0,2\pi]$.}
\label{figDcl}
\end{figure}

The dependence of the diffusion constant for the growth of the variance of the momentum on $K$ is very sensitive, and described in the theoretical result [Eq.~(\ref{Dcl})], and fails around the period 1 accelerator mode intervals
\be \label{acmdint}
(2\pi n) \le K \le \sqrt{(2\pi n)^2 +16 },
\ee
$n$ any positive integer. In these intervals for the accelerator modes $n=1$ we have two \textit{stable fixed points} located at $p=0,\; \theta = \pi -\theta_0$ and $p=0,\; \theta = \pi +\theta_0$, where  $\theta_0 = \arcsin (2\pi/K)$. There are two \textit{unstable fixed points} at $p=0,\; \theta = \theta_0$ and $p=0,\; \theta = 2\pi - \theta_0$. For example, in the case $K=6.5$ we have $\theta_0 \approx 1.31179$. Moreover, as the diffusion might even be anomalous, we have recalculated the \textit{effective} diffusion constant $D_{\rm eff}=\langle(\Delta P)^2\rangle/n$ numerically, which in general is not equal to the $D_{\mu}$ defined in Eq.~(\ref{varp}). In Fig.~\ref{figDcl} we show the $D_{\rm eff}$ for the standard map as a function of $K$ for three discrete times $n$, i.e., the number of the iterations of the standard map, $n=1000$ (lower red dashed line), $n=5000$ (intermediate blue solid) and $n=10000$ (upper black dot-dashed). In the background we have plotted the theoretical diffusion constant $D_1$ taking into account only the normal diffusion (gray dotted line) [Eq.~(\ref{Dcl})]. The presence of  accelerator modes at certain intervals of $K$ (and the sticky objects around) generates anomalous diffusion which is rendered by peaks. Here we used $\approx$ 100000 ($314 \times 314$) initial conditions uniformly distributed in a grid on the entire phase space $[0,2\pi]\times[0,2\pi]$. We see that the dotted theoretical curve stemming from Eq.~(\ref{Dcl}) describes the diffusion constant well outside the accelerator mode intervals. In general, however, the diffusion might be non-normal, described in Eq.~(\ref{varp}). There are also accelerator modes of higher period (2,3,4...) observed and examined below.

\begin{figure}\centering
\includegraphics[width=\columnwidth]{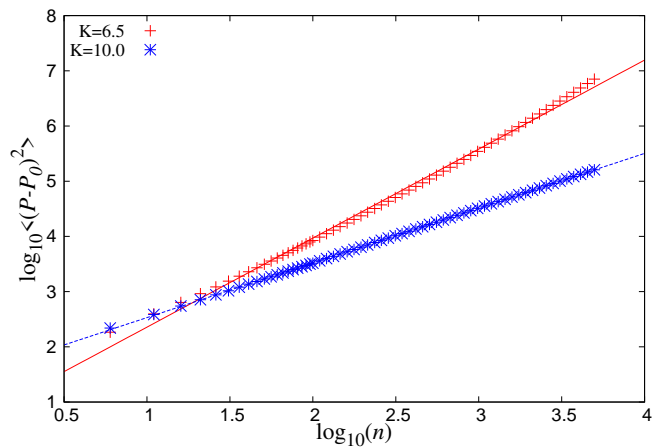}
\caption{(Color online) The variance of the momentum $P$ in the standard map [Eq.~(\ref{SM2})] with $K=6.5$ (red crosses) where small islands and accelerator mode of period 1 are present and $K=10.0$ (blue stars) where the phase space is fully chaotic for the same initial conditions as in Fig.~\ref{figDcl} as a function of the discrete time $n$ (number of iterations), in log-log representation. The two slopes associated with different types of diffusion are $\mu(K=6.5)=1.61252$ (dotted), $\mu(K=10.0)=0.991334$ (solid) with standard deviation errors $\pm$0.01271 (0.7881$\%$) and $\pm$0.0009537 (0.0962$\%$) respectively.}
\label{figDcln}
\end{figure}

\begin{figure*}\centering
\includegraphics[width=15cm,height=8cm]{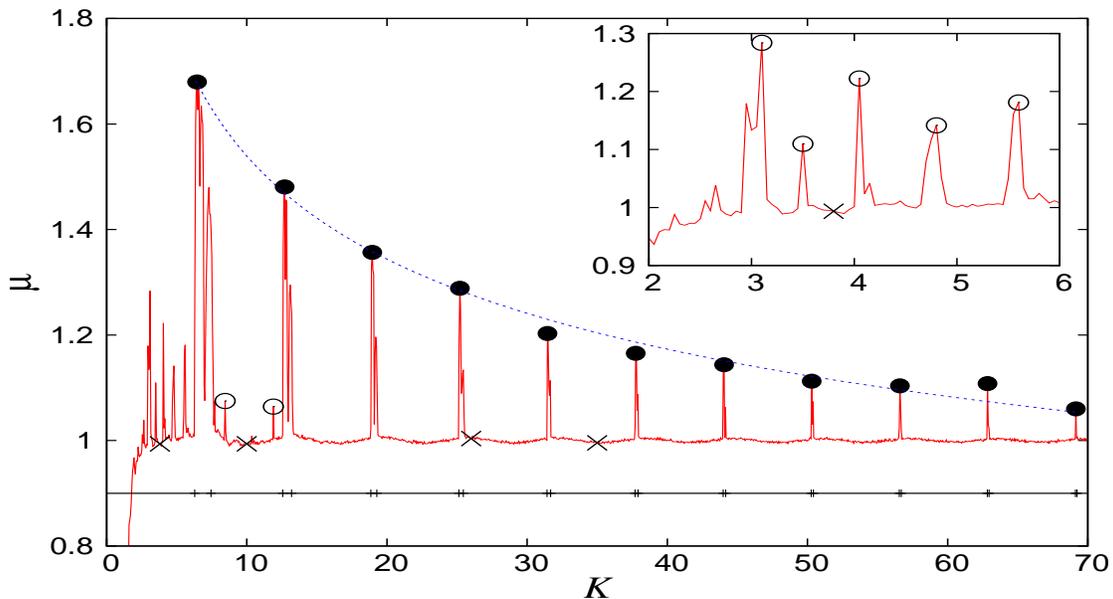}
\caption{(Color online) The diffusion exponent $\mu$ as a function of $K$  after $n=5000$ iterations and for $\approx$100000 $(314\times 314)$ initial conditions on the plane $(\theta,P)=(0,2\pi)$. The intervals on the black horizontal line $\mu=0.9$ indicate the intervals of stable accelerator modes of period 1 [Eq.~(\ref{acmdint})]. All intervals of $K$ with exponent $\mu \approx 1$ are associated with normal diffusion processes. The large peaks (appearing mainly for $K > 2\pi$ marked with full black circles) reflect the anomalous diffusion accelerator modes (mainly of period 1). The smaller peaks for $K < 2\pi$ (more clearly presented in the inset panel) originate by accelerator modes of higher period together with those for $2\pi < K < 4\pi$ marked with empty circle and a few typical examples close to those peaks [marked with the symbol ($\times$)], for which the diffusion is normal, are studied thoroughly later on. The blue dotted line corresponds to the power law which describes the decay of the exponent $\mu$ of the main peaks' amplitude due to accelerator modes of period 1 (see text for more details).}
\label{figKvsmu}
\end{figure*}

In Fig.~\ref{figDcln} we show the variance of the momentum $P$ in the standard map [Eq.~(\ref{SM2})] with $K=6.5$ (red crosses) where small islands and accelerator mode of period 1 are present and $K=10.0$ (blue stars) where the phase space is fully chaotic for the same initial conditions as in Fig.~\ref{figDcl} as a function of the discrete time $n$ (number of iterations), in log-log representation. The two slopes associated with different types of diffusion are $\mu(K=6.5)=1.61252$ (dotted), $\mu(K=10.0)=0.991334$ (solid) with standard deviation errors $\pm$0.01271 (0.7881$\%$) and $\pm$0.0009537 (0.0962$\%$) respectively.
In this manner we have calculated the diffusion exponent $\mu$ for all $K$
on the interval $K\in[K_{cr},70]$ and the result is shown in Fig.~\ref{figKvsmu}. We show the diffusion exponent $\mu$ as a function of $K$ after $n=5000$ iterations, using a fine grid of $314\times 314~(\approx 100000)$ initial conditions on the plane $(\theta,P)=(0,2\pi)$. The $\mu$ is calculated by the slopes, of the lines of the variance of the momentum $P$ as a function of iterations, as it is described in [Eq.~(\ref{varp})] and for a grid of cells on the entire phase space. The intervals on the black horizontal line $\mu=0.9$ indicate the intervals of stable accelerator modes of period 1 [Eq.~(\ref{acmdint})]. All intervals of $K$ with exponent $\mu \approx 1$ are associated with normal diffusion processes. The large peaks (appearing mainly for $K > 2\pi$ marked with full black circles) reflect the anomalous diffusion due to accelerator modes [of period 1, being located inside the intervals predicted by the Eq.~(\ref{acmdint})]. However, there is a number of relatively smaller peaks for $K < 2\pi$ (more clearly presented in the inset panel of Fig.~\ref{figKvsmu}), whose origin is accelerator modes of higher period as we will see later, and also for $2\pi < K < 4\pi$, both these sets are marked with empty circle. With the symbol ($\times$) we mark few typical examples, close to those peaks, for which the diffusion is normal and are also studied in detail in this section.

All the large peaks for $K > 2\pi$, marked with full black circles in Fig.~\ref{figKvsmu}, correspond to regimes with accelerator modes of period 1 and they decrease monotonically as a power law
\be
f(x)=a x^b,
\ee
where $a=2.41645$ and $b=-0.195896$ [blue dotted line in Fig.~\ref{figKvsmu}, with asymptotic standard error $\pm$0.04294 (1.777$\%$) and $\pm$0.00537 (2.741$\%$) respectively] indicating that for $K>70$ their effect decreases significantly. On the other hand, the size of the successive accelerator modes of period 1 intervals decays with a power law defined simply and analytically by the Eq.~(\ref{acmdint}).
In order to understand the effect of the presence of accelerator modes in the diffusion and transport properties of the phase space in the standard map, we first picked an, as much as possible, representative sample of $K$-values. In more detail, we included in our test-cases all the $K$-values which correspond to all the main peaks appearing in Fig.~\ref{figKvsmu} with $K > 2\pi$ together with a few cases from the `plateaus' of this curve. Furthermore, we took into account the peaks occurring for $1 \lesssim K \lesssim 2\pi$ (see the empty black circles in the inset zoom in Fig.~\ref{figKvsmu}) which are associated with accelerator modes of higher periodicity, as it will be seen thereupon. The case with $K=3.8$, whose $\mu$ value is $\approx 1$, is chosen for comparison reasons from the plateau and as it turns out has no accelerator modes in its phase space causing anomalous diffusion. Here we should stress that we repeated the same procedure for larger number of iterations $n=10000$ and it turns out that the exponent $\mu$ has well converged to the values shown in Fig.~\ref{figKvsmu}.

The distribution of the momenta in the case of normal diffusion is found to be
perfect Gaussian, whilst for anomalous diffusion a strong departure from
the Gaussian distribution is observed, being well fitted by a stable L\'{e}vy distribution, characterized by the parameter $\alpha\in [0,2]$. The details are given in reference \cite{MR2014}. For each one of the $K$-values of the nonlinearity kick parameter of Fig.~\ref{figKvsmu}, we have performed a thorough study by calculating and comparing the following quantities

(a) The index of stability $\alpha$-parameter of the L\'evy stable distribution.\

(b) The diffusion exponent $\mu$ as described in Eq.~(\ref{varp}).\

In the case of normal diffusion (Gaussian statistics) for the above quantities, one expects to find $\alpha=2$ for the L\'evy stable distribution and diffusion exponent $\mu=1$, while in the general case we find other values. We have calculated the $\mu$ exponent emerging from a small box/ensemble of initial conditions in the phase space and thus produced the $\mu$-landscape in the phase space of the SM. Furthermore, we have also employed the GALI-method
\cite{SBA:2007,SM2014} to calculate the GALI-index in order to identify the regular and chaotic regions in the phase space of the SM, which also quantifies the degree of chaos (indirectly). The GALI-landscape and the $\mu$-landscape are found to correspond very well to each other, where it seems that the $\mu$-plot contains more information than GALI-plot.

The conclusion of this section is that the SM exhibits normal diffusion for most of the $K$ values on the interval $[K_{cr},70]$, except for the accelerator mode intervals where anomalous diffusion is observed, with the exponent $\mu$ typically being larger than 1. Using these plots and the described methodology we found that the case $K=10$ is the closest to full chaos (no regular islands present) and exhibits the normal diffusion for all initial conditions in the phase space of the SM.

\section{The  quantum kicked rotator and the Izrailev model }
\label{sec2}

The quantum kicked rotator (QKR) is the quantized version of Eq.~(\ref{KR}),
namely
\be \label{QKR}
\hat{H} =-\frac{\hbar^2}{2I} \frac{\partial^2}{\partial\theta^2} +
V_0\, \delta_T(t)\,\cos \theta .
\ee
The Floquet operator $\hat{F}$ acting on the wavefunctions (probability amplitudes) $\psi(\theta)$, $\theta \in[0,2\pi)$, upon each period (of length $T$) can be written as (see e.g. \cite{Stoe}, Chapter 4)
\be \label{Fop}
\hat{F} = \exp \left( -\frac{iV_0}{\hbar} \cos\theta\right)
\exp\left(-\frac{i\hbar T}{2I}\frac{\partial^2}{\partial\theta^2}\right),
\ee
where now we have two {\em dimensionless} quantum control parameters
\be \label{qpar}
k=\frac{V_0}{\hbar}, \;\;\; \tau= \frac{\hbar T}{I},
\ee
which satisfy the relationship $K = k\tau = V_0 T/I$, $K$ being the classical {\em dimensionless} control parameter of Eq.~(\ref{SM2}). By using the angular momentum eigenfunctions
\be \label{Eigenf}
|n\rangle = a_n(\theta) = \frac{1}{\sqrt{2\pi}} \exp (i\,n\,\theta),
\ee
where $n$ is any integer, we find the matrix elements of $\hat{F}$, namely
\begin{flalign} \label{Fmatrix}
F_{m\,n} = \langle m|\hat{F}|n\rangle = \exp\left(-\frac{i\tau}{2} n^2\right) i^{n-m} J_{n-m} (k),
\end{flalign}
where $J_{\nu}(k)$ is the $\nu$-th order Bessel function. For a wavefunction
$\psi (\theta)$ we shall denote its angular momentum component (Fourier
component) by
\begin{flalign} \label{Fourier} \nonumber
u_n = \langle n|\psi\rangle = \int_0^{2\pi} a_n^{*}(\theta) \psi(\theta)\,d\theta= \\ =\frac{1}{\sqrt{2\pi}} \int_0^{2\pi} \psi(\theta) \exp(-in\theta)\,d\theta.
\end{flalign}
The QKR has very complex dynamics and spectral properties. As the phase space is infinite (cylinder), $p\in (-\infty, +\infty), \theta\in[0,2\pi)$, the spectrum of the eigenphases of $\hat{F}$, denoted by $\phi_n$, or the associated quasienergies $\hbar\om_n= \hbar \phi_n/T$, introduced by Zeldovich  \cite{Zel1966}, can be continuous, or discrete. It is quite well understood that for the resonant values of $\tau$
\be \label{Restau}
\tau = \frac{4\pi r}{q},
\ee
$q$ and $r$ being positive integers without common factor, the spectrum is continuous, as rigorously proven by Izrailev and Shepelyansky \cite{IS1979a,IS1979b,IS1980a,IS1980b}, and the dynamics is (asymptotically) ballistic, meaning that starting from an arbitrary initial state the mean value of the momentum $\langle\hat{p}\rangle$ increases {\em linearly} in time, and the energy of the system $E= \langle\hat{p}^2\rangle/(2I)$ grows {\em quadratically} without limits. For the special case $q=r=1$ this can be shown elementary. Such behavior is a purely quantum effect, called the quantum resonance. Also, the regime of quadratic energy growth manifests itself only after very large time, which grows very fast with the value of the integer $q$ from Eq.~(\ref{Restau}), such that for larger $q$ this regime practically cannot be observed.

For generic values of $\tau/(4\pi)$, being irrational number, the spectrum is expected to be discrete but infinite. But the picture is very complicated. Casati and Guarneri \cite{CG1984} have proven that for $\tau/(4\pi)$ sufficiently close to a rational number, there exists a continuous component in the quasienergy spectrum. So, the absence of dynamical localization for such cases is expected as well. Without a rigorous proof, we finally believe that
for all other (``good") irrational values of $\tau/(4\pi)$ we indeed have discrete spectrum and quantum dynamical localization. In such case the quantum dynamics is almost periodic, and because of the effective finiteness of the relevant set of components $u_n$ and of the basis functions involved, just due to the exponential localization (see below), it is even effectively quasiperiodic (effectively there is a finite number of frequencies), and any initial state returns after some recurrence time arbitrarily close to the initial state. Thus the energy cannot grow indefinitely.

The asymptotic localized eigenstates are {\em exponentially localized}. The (dimensionless) theoretical localization length in the space of the angular momentum quantum numbers is given below, and is equal (after introducing some numerical correction factor $\alpha_{\mu}$) to the dimensionless localization time $t_{\rm loc}$  [Eq.~(\ref{finallinf})]. We denote it unlike in reference \cite{Izr1990} and \cite{MR2013} by $\linf$.  Therefore, an exponentially localized eigenfunction centered at $m$ in the angular momentum space [Eq.~(\ref{Eigenf})] has the following form
\be \label{exploc} |u_n|^2 \approx \frac{1}{\linf} \exp\left(-\frac{2|m-n|}{\linf}\right),
\ee
where $u_n$ is the probability amplitude [Eq.~(\ref{Fourier})] of the localized wavefunction $\psi(\theta)$. The argument leading to $t_{\rm loc}$ in Eq.~(\ref{finallinf}) originates from the observation of the dynamical localization by Casati \textit{et al.} \cite{CCFI79}, and in particular from \cite{CIS1981}, and is well explained in \cite{Stoe}, in case of normal diffusion $\mu=1$, whilst for general $\mu$ we gave a theoretical argument in \cite{MR2013}. We shall denote $\sigma = 2/\linf$, and will
later on determine the $\sigma$'s directly from the individual numerically calculated eigenstate.

The question arises, where do we see the  phenomena (spectral statistics, namely Brody-like level spacing distribution) analogous  in the quantum chaos of time-independent bound systems with discrete spectrum?  To see these effects the system must have effectively finite dimension, because in the infinite dimensional case we simply observe Poissonian statistics. Truncation of the infinite matrix $F_{mn}$ in Eq.~(\ref{Fmatrix}) in {\em tour de force} is not acceptable, even in the technical case of numerical computations, since after truncation the Floquet operator is no longer unitary.

The only way to obtain a quantum system which shall in this sense correspond to the classical dynamical system [Eqs.~(\ref{KR}), (\ref{SM1}) and (\ref{SM2})] is to introduce a finite $N$-dimensional matrix, which is symmetric unitary, and which in the limit $N\rightarrow\infty$ becomes the infinite dimensional system with the Floquet operator [Eq.~(\ref{Fop})]. The semiclassical limit is $k\rightarrow \infty$ and $\tau\rightarrow 0$, such that $K=k\tau ={\rm constant}$. As it is well known \cite{Izr1990}, for the reasons discussed above, the system behaves very similarly for rational and irrational values of $\tau/(4\pi)$. Such a $N$-dimensional model \cite{Izr1988} will be introduced below.

Following \cite{CIS1981,MR2013} we find that the dimensionless Heisenberg
time, also called break time or localization time, denoted by $t_{\rm loc}$,
in units of kicking period $T$, is equal to the dimensionless
localization length $\linf$ as shown in Eq.~(\ref{tloc=linf}).
\be \label{finallinf}
\linf \approx t_{\rm loc} = \left( \alpha_{\mu} \frac{D_{\mu}(K)}{\tau^2} \right)^{\frac{1}{2-\mu}}.
\ee
where $\alpha_{\mu}$ is a numerical constant to be determined empirically,
and in case of normal diffusion $\mu=1$ is close to $1/2$. Since this
semiclassical approach and derivation is quite important, we repeat the
arguments given in \cite{MR2013}.

The generalized diffusion process of the standard map [Eq.~(\ref{SM2})] is defined by Eq.~(\ref{varp}). As the real physical angular momentum $p$ and $P$ are connected by $P=pT/I$ we have for the variance of $p$ the following equation
\be \label{varpreal}
\langle (\Delta p)^2\rangle = \frac{I^2}{T^2} D_{\mu} n^{\mu}.
\ee
(Please note that in ref. \cite{MR2013} there is a typing error in formula (13), where $T$ and $I$ must be switched.) Now we argue as follows: The general wisdom (golden rule) in quantum chaos is that the quantum diffusion follows the classical diffusion up to the Heisenberg time (or break time, or localization time), defined as
\be \label{tH}
t_H = \frac{2\pi\hbar}{\Delta E},
\ee
where $\Delta E$ is the mean energy level spacing. In our case we have the quasienergies and $\Delta E= \hbar \Delta\omega$, where $\Delta \omega = \Delta \phi/T$, and $\Delta \phi$ is the mean spacing of the eigenphases. This might be estimated at the first sight as $\Delta \phi = 2\pi/N$, but this is an underestimate, as effectively we shall have due to the localization only $\linf$ levels on the interval $[0,2\pi)$. Therefore $\Delta \phi = 2\pi/\linf$ and we find
\be \label{tH2}
t_H = \frac{2\pi T}{\Delta \phi} = T\linf.
\ee
Since $T$ is the period of kicking, and $t_H$ is the real physical continuous time, we get the result that the discrete time [number of iterations of Eq.~(\ref{SM2}) at which the quantum diffusion stops], the localization time $t_{\rm loc}$ is indeed equal to the localization length in momentum space, i.e.
\be \label{tloc=linf}
t_{\rm loc} \approx \linf.
\ee
Since our derivation is not rigorous, we use the approximation symbol rather than equality, in particular as the definition depends linearly on the definition of the Heisenberg time. Now the final step: By inspection of the dynamics of the Floquet quantal system [Eqs.~(\ref{QKR}),(\ref{Fop})] one can see (see also the derivation in the St\"ockmann's book \cite{Stoe}) that the value of the variance of the angular momentum at the point of stopping the diffusion $\langle (\Delta p)^2\rangle$ is proportional to $\hbar^2 \linf^2$, and to achieve equality we introduce a dimensionless numerical (empirical,correction) factor $\alpha_{\mu}$ by writing $\langle (\Delta p)^2\rangle = \hbar^2 \linf^2 /\alpha_{\mu}$, which on the other hand must be equal just to the classical value at stopping time $t_{\rm loc}$, namely equal to $(I/T)^2D_{\mu} \linf^{\mu}$. From this it follows immediately Eq.~(\ref{finallinf}).
The numerical constant $\alpha_{\mu}$ is found empirically by numerical calculations, for instance  in the literature the case $K=5$ with $\mu=1$ is found to be $\alpha_1=0.5$ (however, we find numerically $\alpha_1=0.45$,  taking into account Eq.~(\ref{finallinf}) when studying the model's localization properties). Thus, we have the theoretical formula for the localization length in the case of generalized  classical diffusion [Eqs.~(\ref{varp}),(\ref{varpreal})], which we use in defining the scaling parameter $\Lambda$ below. 

The motion of the QKR [Eq.~(\ref{QKR})] after one period $T$ of the $\psi$ wavefunction can be described also by the following symmetrized Floquet mapping, describing the evolution of the kicked rotator from the middle of a free rotation over a kick to the middle of the next free rotation, as follows
\begin{flalign} \label{Uoper}
& \psi(\theta,t+T) = \hat{U}\psi(\theta,t), \\ \nonumber
& \hat{U} = \exp \left ( i \frac{T\hbar}{4I}\frac{\partial^2}{\partial \theta^2} \right )\exp \left (-i\frac{V_0}{\hbar} \cos \theta \right)\exp \left (i \frac{T\hbar}{4I}\frac{\partial^2}{\partial \theta^2} \right).
\end{flalign}
Thus, the $\psi(\theta,t)$ function is determined in the middle of the rotation, between two successive kicks. The evolution operator $\hat{U}$ of the system corresponds to one period. Due to the instant action of the kick, this evolution can be written as the product of three non-commuting unitary operators, the first and third of which correspond to the free rotation during half a period $\hat{G}(\tau/2)=\exp \left (i\frac{\tau}{4}\frac{\partial^2}{\partial \theta^2} \right)$, $ \tau \equiv \hbar T/I$, while the second $\hat{B}(k)=\exp(-ik\cos \theta)$, $ k \equiv V_0/\hbar$, describes the kick. Like before, we have only two dimensionless parameters, namely $\tau$ and $k$, and  $K=k \tau=V_0 T/I$. In the case $K\equiv k \tau \gg 1$ the motion is well known to be strongly chaotic,
for $K=10$ certainly without any regular islands of stability, as mentioned,
and also there are no accelerator modes, so that the diffusion is normal
($\mu=1$). We have carefully checked that the case $K=10$ is the closest
to the normal diffusion $\mu=1$  for all $K\in [K_{cr},70]$. The transition to classical mechanics is described by the limit $k \rightarrow \infty$, $\tau \rightarrow 0$ while $K=\rm{const}$. We shall consider the regimes on the
interval $3\le k\le 20$, but will concentrate mostly on the semiclassical regime $k\ge K$, where $\tau \le 1$.

In order to study how the localization affects the statistical properties of the quasienergy spectra, we use the model's representation in the momentum space with a finite number $N$ of levels \cite{Izr1988,Izr1990,Izr1986,Izr1987,Izr1989}, which we refer to as Izrailev model
\begin{flalign} \label{u_repres}
u_n(t+T) = \sum_{m=1}^{N} U_{nm}u_m(t), \ n,m=1,2,...,N \enspace .
\end{flalign}
The finite symmetric unitary matrix $U_{nm}$ determines the evolution of an $N$-dimensional vector, namely the Fourier transform $u_n(t)$ of $\psi(\theta,t)$, and is composed in the following way
\be \label{Unm}
    U_{nm}=\sum_{n'm'}G_{nm'}B_{n'm'}G_{n'm},
\ee
where $G_{ll'}=\exp \left (i\tau l^2/4 \right )\delta_{ll'}$ is a diagonal matrix corresponding to free rotation during a half period $T/2$, and  the matrix $B_{n'm'}$ describing the one kick has the following form
\begin{flalign} \label{Bnmoper}\nonumber
  & B_{n'm'}= \frac{1}{2N+1}\times \\ \nonumber
  & \sum_{l=1}^{2N+1} \left \{ \cos \left [ \left (n'-m' \right ) \frac{2 \pi l}{2N+1}\right ] - \cos \left [(n'+m')\frac{2 \pi l}{2N+1} \right ] \right \} \\
  & \times  \exp \left [-ik\cos \left (\frac{2 \pi l}{2N+1}\right ) \right ].
\end{flalign}
The Izrailev model in Eqs.~(\ref{u_repres}-\ref{Bnmoper}) with a finite number of states is considered as the quantum analogue of the classical standard mapping on the torus with closed momentum $p$ and phase $\theta$, where $U_{nm}$ describes only the odd states of the systems, i.e. $\psi(\theta)=-\psi(-\theta)$, provided we have the case of the quantum resonance, namely $\tau =4\pi r/(2N+1)$, where $r$ is a positive integer, as in Eq.~(\ref{Restau}). The matrix (\ref{Bnmoper}) is obtained by starting the derivation from the odd-parity basis of $\sin(n\theta)$ rather than the general angular momentum basis $\exp(in\theta)$.

Nevertheless, we shall use this model for any value of $\tau$ and $k$, as a model which in the resonant and in the generic case (irrational $\tau/(4\pi)$) corresponds to the classical kicked rotator, and in the limit $N\rightarrow \infty$ approaches the infinite dimensional model [Eq.~(\ref{Uoper})], restricted to the symmetry class of the odd eigenfunctions. It is of course just one of the possible discrete approximations to the continuous infinite dimensional model.

The difference of behavior between the generic case and the quantum resonance shows up only at very large times, which grow fast with $(2N+1)$, as explained
above.  It turns out that also the eigenfunctions and the spectra of the eigenphases at finite dimension $N$ of the matrices that we consider do not show any significant differences in structural behavior for the rational or irrational $\tau/(4\pi)$, which we have carefully checked. Indeed, although the eigenfunctions and the spectrum of the eigenphases exhibit {\em sensitive dependence on the parameters} $\tau$ and $k$, their statistical properties are stable against the small changes of $\tau$ and $k$. This is an advantage, as instead of using very large single matrices for the statistical analysis, we can take a large ensemble of smaller matrices for values of $\tau$ and $k$ around some central value of $\tau=\tau_0$ and $k=k_0$, which greatly facilitates the numerical calculations  and improves the statistical significance of our empirical results. Therefore our approach is physically meaningful. Similar approach was undertaken by Izrailev (see \cite{Izr1990}
and references therein). In Fig.~1 of paper \cite{MR2013} we show the examples of strongly exponentially localized eigenstates by plotting the natural logarithm of the probabilities  $w_n=|u_n|^2$ versus the momentum quantum number $n$, for two different matrix dimensions $N$. By calculating the localization length $\linf$ from the slopes $\sigma=2/\linf$ of these eigenfunctions using Eq.~(\ref{exploc}) we can get the first quantitative empirical localization measure to be discussed and used later on. The new finding of our paper \cite{MR2015} is that $\sigma$ has a distribution, which is close to the Gaussian (but cannot be exactly that, because $\sigma$ is a positive definite quantity). It does not depend on $N$ and survives the limit $N\rightarrow \infty$. Therefore also $\linf$ has a distribution whose variance does not vanish in the limit $N\rightarrow \infty$.

Following \cite{MR2013} and \cite{Izr1990} we introduce another measure of localization. For each $N$-dimensional eigenvector of the matrix $U_{nm}$ the information entropy is
\be  \label{infoentr}
 \mathscr{H}_N(u_1,...,u_N) = -\sum_{n=1}^{N}w_n \ln w_n,
\ee
where $w_n = |u_n|^2$, and $\sum_n |u_n|^2 = 1$.

In case of the random matrix theory being applicable to our system [Eqs.~(\ref{Uoper}) and (\ref{u_repres}-\ref{Bnmoper})], namely the COE (or GOE), due to the isotropic distribution of the eigenvectors of a COE of random matrices, we have the probability density function of $|u_n|$ on the interval $[0,1]$,
\begin{flalign} \label{GOEeigvec}
w_N(|u_n|) = \frac{2\Gamma(N/2)}{\sqrt{\pi} \Gamma((N-1)/2) } (1-|u_n|^2)^{(N-3)/2}.
\end{flalign}
It is easy to show that in the limit $N\rightarrow \infty$ this becomes a Gaussian distribution
\be \label{Gaussianeigvec}
w_N(|u_n|) = \sqrt{\frac{2N}{\pi}} \exp \left(-\frac{N|u_n|^2}{2}
\right),
\ee
and the corresponding information entropy [Eq.~(\ref{infoentr})] is equal to
\begin{flalign} \label{H_GOE}
 \mathscr{H}_{N}^{GOE}=\psi \left (\frac{1}{2}N+1 \right )-\psi \left (\frac{3}{2} \right )\simeq \ln \left (\frac{1}{2}Na \right )+O(1/N),
\end{flalign}
where $a=\frac{4}{\exp(2-\gamma)}\approx 0.96$, while $\psi$ is the digamma function and $\gamma$ the Euler constant ($\simeq 0.57721...$). For a uniform distribution over $M$ states $w_n=1/M$ we get $\mathscr{H}_{N} \approx \log M$, and thus $M \approx \exp (\mathscr{H}_{N})$. Thus, we get the insight that the correct measure of localization must be proportional to $\exp (\mathscr{H}_{N})$, but properly normalized, such that in case of extendedness (GOE/COE) it is equal to $N$.

\begin{figure*} \centering 
\center
\includegraphics[width=15cm]{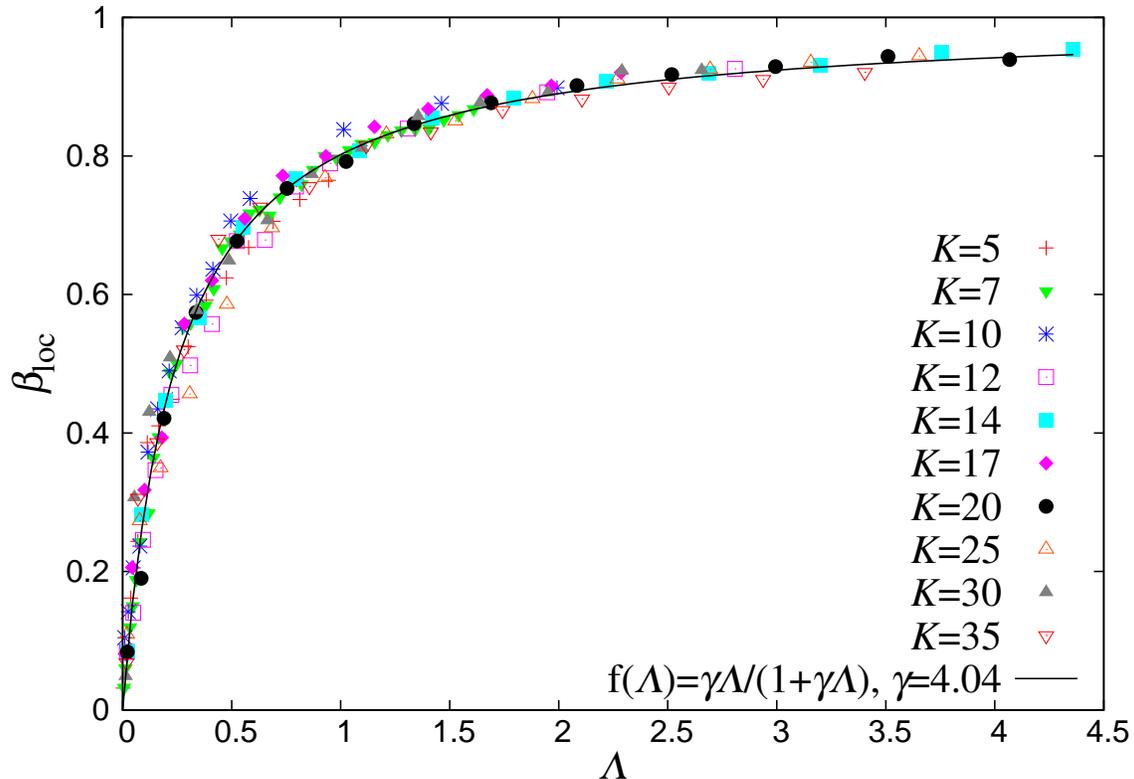}
\caption{ [Color online] The parameter $\blo$ vs. $\La$ for $161\times398$ elements for various values of $K$ and for a wide range of $k$ values,  where the scaling law [Eq.~(\ref{bla})] is shown with the black line. }
\label{fig4}
\end{figure*}

Therefore the {\it entropy localization length} $l_H$ is defined as \be\label{lh:eq}
 l_H=N \exp \left (\mathscr{H}_{N}-\mathscr{H}_{N}^{GOE} \right ).
\ee
Indeed, for entirely extended eigenstates $l_H=N$. Thus, $l_H$ can be calculated for every eigenstate individually. However, all eigenstates,
while being quite different in detail, are exponentially localized,
and thus statistically very similar. Therefore, in order to minimize the fluctuations one uses the {\it mean localization length} $d\equiv \langle l_H \rangle$, which is computed by averaging the entropy over all eigenvectors of the same matrix (or even over an ensemble of similar matrices)
\be\label{d:eq}
 d \equiv \langle l_H \rangle = N \exp \left (\langle \mathscr{H}_{N} \rangle-\mathscr{H}_{N}^{GOE} \right ).
\ee
The {\it localization parameter} $\blo$ is then defined as
\be\label{beta_loc:eq}
    \blo=\frac{d}{N}\equiv \frac{\langle l_H\rangle}{N}.
\ee
The parameter that determines the transition from weak to strong quantum chaos is neither the strength parameter $k$ nor the localization length $\linf$, but the ratio of the localization length $\linf$ to the size $N$ of the system in momentum $p$
\be\label{MLL:Lamda}
  \La=\frac{\linf}{N} = \frac{1}{N}
\left( \frac{\alpha_{\mu} D_{\mu} (K)}{\tau^2}\right)^{\frac{1}{2-\mu}} ,
\ee
where $\linf \approx t_{\rm loc}$, the theoretical localization length
Eq.~(\ref{finallinf}), was derived in \cite{MR2013}. $\La$ is the scaling parameter of the system.  The relationship of $\La$ to $\blo$ is discussed e.g. in \cite{CGIS1990,Izr1990} and further developed in section VII of \cite{MR2013}. Here we just summarize by showing the empirical scaling property of $\La$ versus $\blo$ in Fig.~\ref{fig4}, where the approximate analytical description is given by the function
\be \label{bla}
    \blo = \frac{\gamma \La}{1 +\gamma \La},\;\;\; \gamma=4.04,
\ee
which is similar to the scaling law in in \cite{CGIS1990}, but not the same.
Namely, the value $\gamma=4.04$ differs somewhat from  $\gamma\approx 3.2$ in
\cite{CGIS1990}, where $\blo$ is plotted versus $x\approx 4\La$.

In producing this plot we have used Eqs.~(\ref{finallinf}),(\ref{MLL:Lamda}) for $\La$, which is just a rough estimate. Indeed, as we shall show below, following \cite{MR2015}, $\La$ is not a number in a given system, at fixed $K$, $k$ and $N$, but has a distribution, whose reciprocal is approximately Gaussian distributed, and  Eq.~(\ref{finallinf}) is just a rough estimate of the mean value of $\La$. Therefore we should not be surprised any more to see large fluctuations in the scaling law of Fig. \ref{fig4}, an observation entirely unexplained so far, but clarified in \cite{MR2015}. The statistical properties of the localization measure will be discussed below.

\begin{figure*} \centering
\center
\includegraphics[width=15cm]{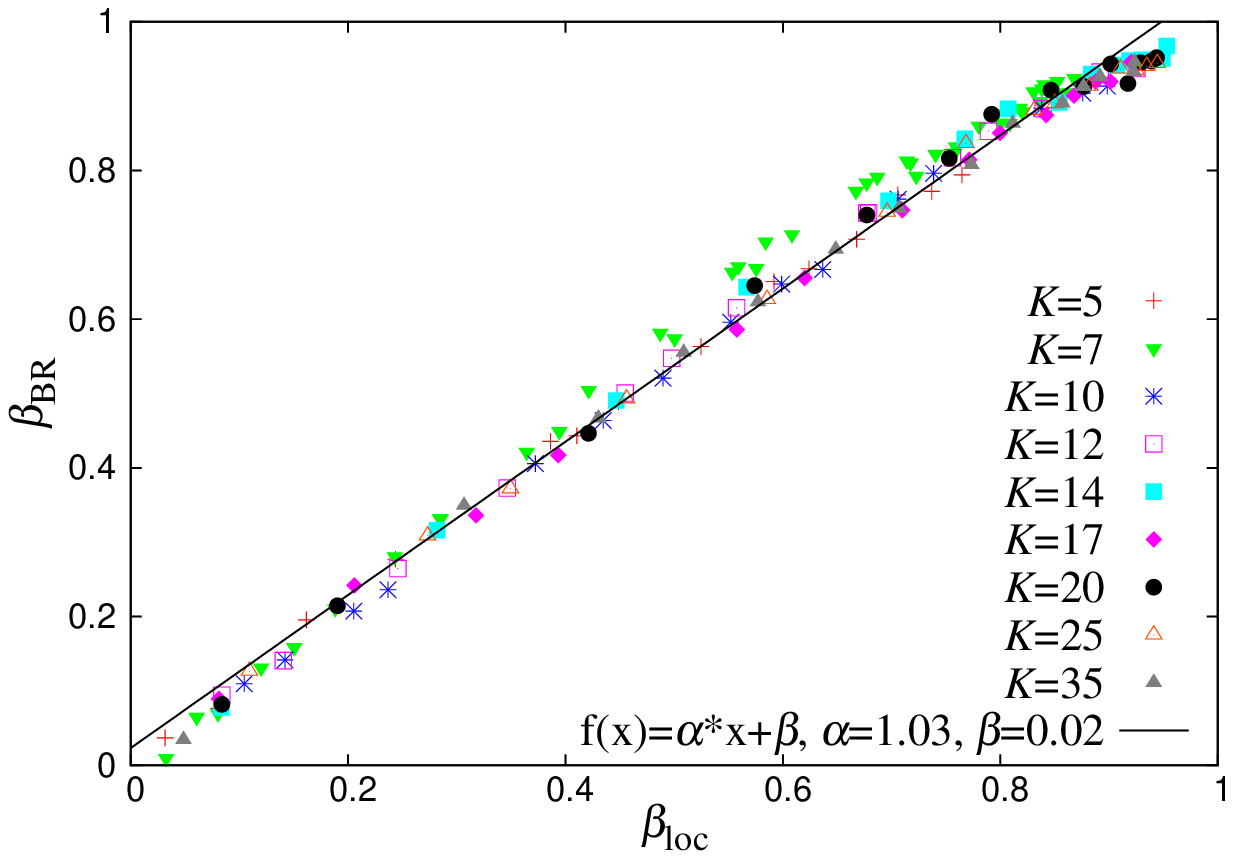}
\caption{[Color online] The fit parameter $\bro$ as a function of $\blo$ for $161\times398$ elements for various values of $K$ and for a wide range of $k$ values. The best fitting straight line is very close to identity.}
\label{fig5}
\end{figure*}

In closing this section we just mention the important finding \cite{MR2013} that the level spacing distribution $P(S)$ (of the quasienergies) is very well described by the Brody distribution \cite{Bro1973,Bro1981}
\be \label{BrodyDistrib}
    P_{\rm BR}(S)=C_1 S^\beta \exp \left (-C_2 S^{\beta+1} \right ),
\ee
where the two parameters $C_1$ and $C_2$ are determined by the two generic normalization conditions that must be obeyed by any $P(S)$,
\be\label{NormCond}
    \int_0^\infty P(S) dS=1, \quad \langle S \rangle=\int_0^\infty S P(S) dS=1,
\ee
thus with $\langle S \rangle=1$ being the mean distance between neighboring levels (after unfolding). Hence
\be\label{C1C2}
    C_1=(\beta+1)C_2, \quad C_2=\left[\Gamma\left(\frac{\beta+2}{\beta+1}\right)\right]^{\beta+1},
\ee
where $\Gamma(x)$ denotes the Gamma function. In the strongly localized regime $\beta=0$ we observe Poissonian statistics while in the fully chaotic one $\beta=1$ and the RMT applies. The Brody cumulative level spacing distribution is
\be \label{BrodyCum}
    W_{\rm BR}(S)=1-\exp(-C_2 S^{\beta+1}).
\ee
In Fig.~\ref{fig5} we show the scaling of the Brody spectral parameter $\bro$ vs. the localization parameter $\blo$. They are linearly related almost like identity, but we also observe rather large fluctuations, probably due to the fact that $\bro$ is difficult to calculate accurately enough with small matrices.

\section{The distribution of the localization measures}
\label{sec3}

\begin{figure}
\center
\includegraphics[width=7.5cm]{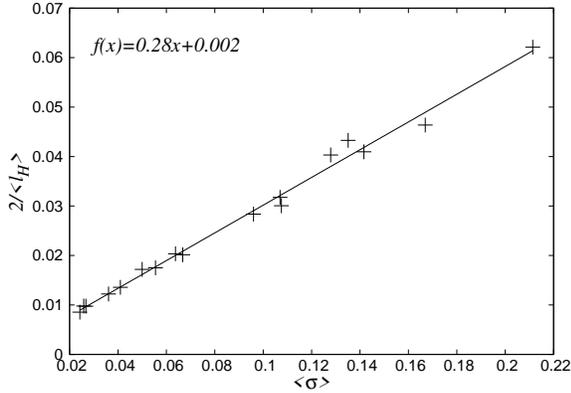}
\caption{We show $\langle\sigma \rangle$ versus $2/\langle l_H \rangle$
for matrices of dimension $N=3000$, for 7 nearby values of $k$, namely $k=k_0 \pm j\delta k$, where $j=0,1,2,3$ and $\delta k=0.00125$, for $k_0=3,4,5,\dots,19$. The two empirical localization measures are clearly well defined, linearly related and thus equivalent.}
\label{fig6}
\end{figure}

In this section we follow our recent paper \cite{MR2015} and present the results about the distribution of the localization measures. Here we restrict our analysis exclusively to the case $K=10$, as this case is closest to the normal diffusion regime $\mu=1$, as explained in Sec.~\ref{sec1}. First we demonstrate that the localization measures  $\linf=2/\sigma$ and $l_H$  are very well defined, linearly related and thus equivalent. In Fig.~\ref{fig6} we show this in the  diagram  of the mean $\langle \sigma \rangle$ versus $2/\langle l_H \rangle$, where both averagings are over all eigenfunctions for  matrices of dimension $N=3000$, for 7 nearby values of $k$ around $k_0$, namely $k=k_0 \pm j\delta k$, where $j=0,1,2,3$ and $\delta k=0.00125$, for $k_0=3,4,5,\dots,19$.

In the next Fig.~\ref{fig7} we show the relationship of the theoretical $\linf$ in Eq.~(\ref{finallinf}) and the mean value  of the empirical $2/\langle \sigma\rangle$  for  $k_0=3,4,5,...,19$. It is clearly seen in Fig.~\ref{fig7}(a) that there are strong fluctuations which we attribute to the fact that $2/\sigma$ has a certain distribution with nonvanishing variance, to be presented and described below, and that the theory of $\linf$ resulting in Eq.~(\ref{finallinf}) is too simple, as it corresponds only roughly to the value of $2/\langle\sigma\rangle$. On the other hand, in Fig.~\ref{fig7}(b) we see again that the two empirical localization measures are exactly linearly related. We should mention that in the cases of larger $k > 19$ the slopes $\sigma$ are so small, and the localization too weak, that we cannot get reliable results. Therefore we limit ourselves to the interval $3\le k\le19$.

\begin{figure}
\center
\includegraphics[width=7.5cm]{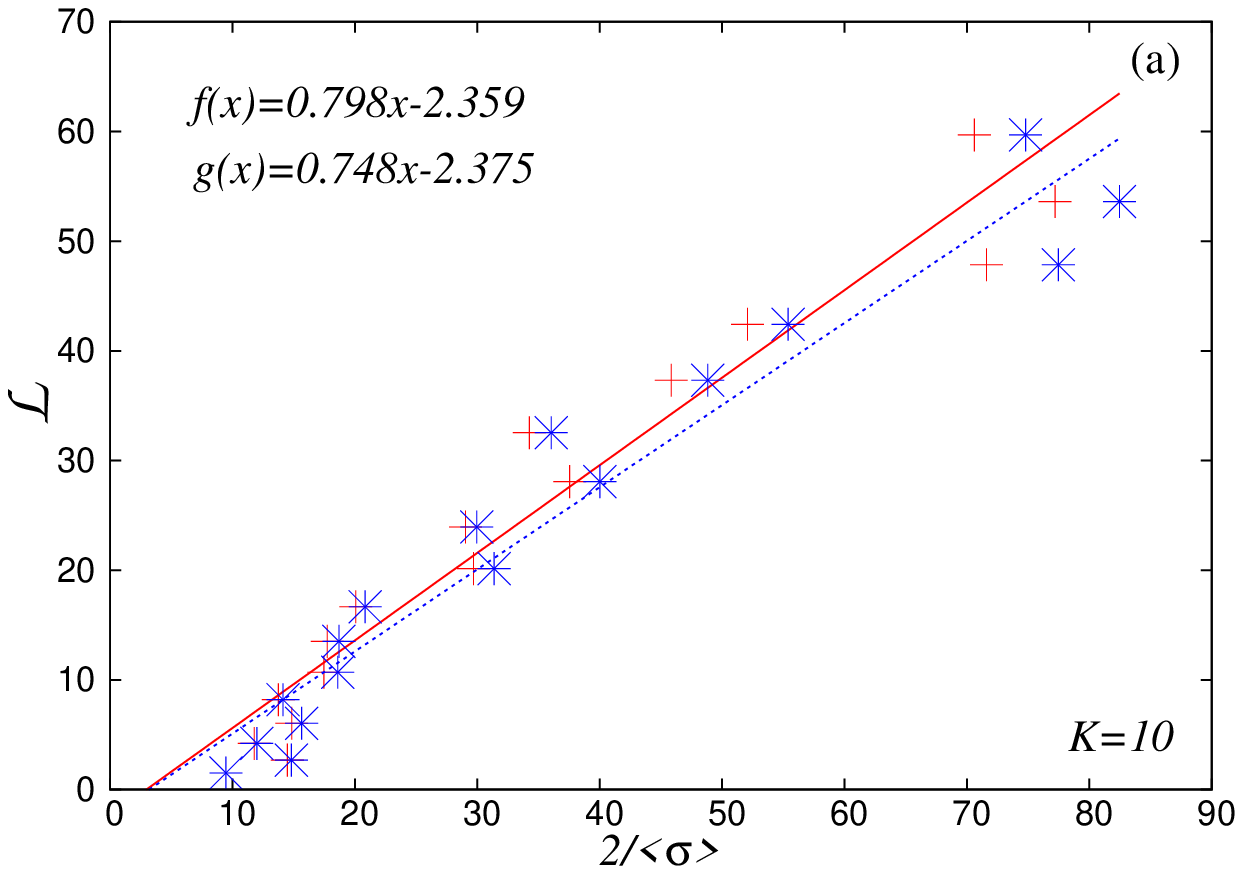}
\includegraphics[width=7.5cm]{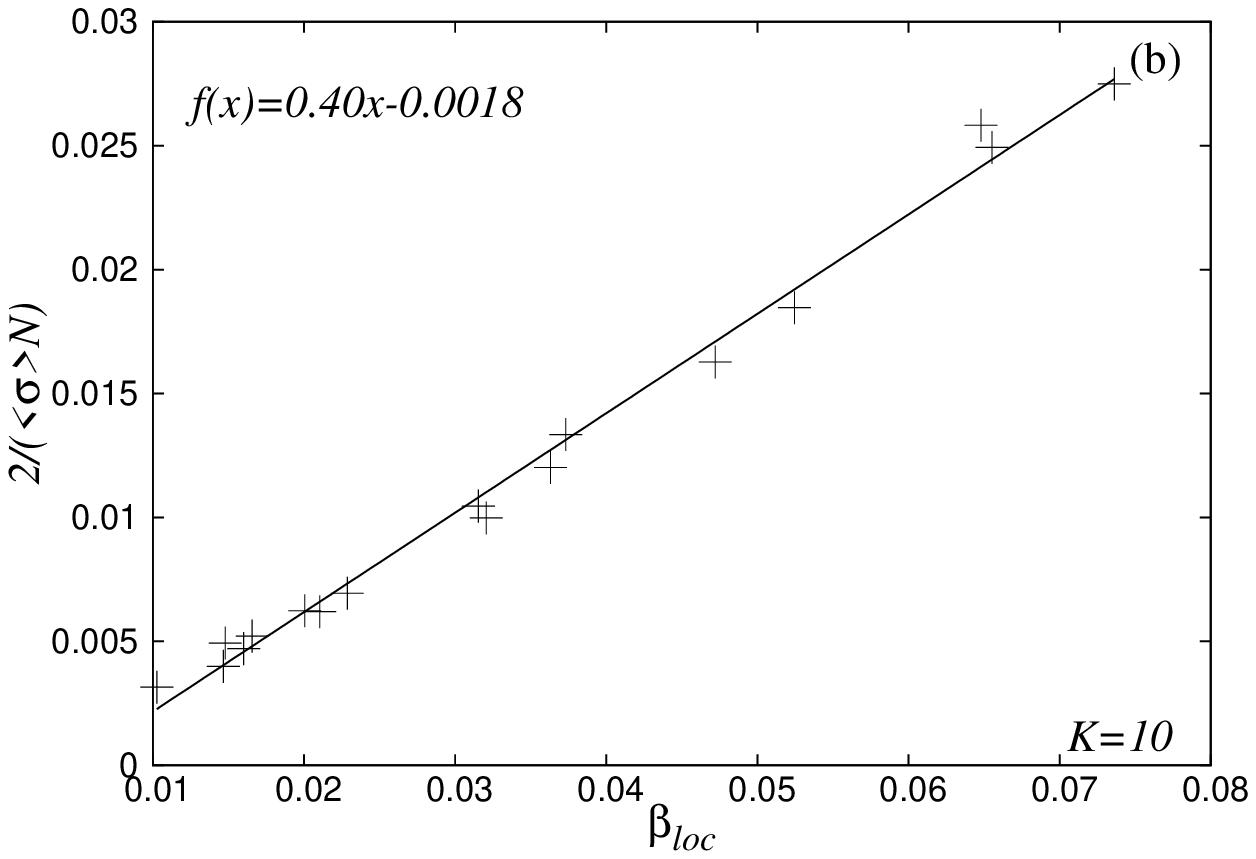}
\caption{[Color online] (a) We show $\linf$ versus $2/\langle \sigma\rangle$
for matrices of dimension $N=1000$ (crosses and solid fit line) and for matrices of dimension $N=3000$ (stars and dashed fit line), for 7 nearby values of $k$, namely $k=k_0 \pm j\delta k$, where $j=0,1,2,3$ and $\delta k=0.00125$, for $k_0=3,4,5,...,19$.  (b) We plot the mean value of $2/(N\langle\sigma\rangle)$ versus $\blo$ for $k_0=3,4,5,\dots,19$ and
7 matrices of dimension $N=3000$ with $k=k_0 \pm j\delta k$,
where $j=0,1,2,3$ and the step size $\delta k=0.00125$.}
\label{fig7}
\end{figure}

\begin{figure}
\center
\includegraphics[width=7.5cm]{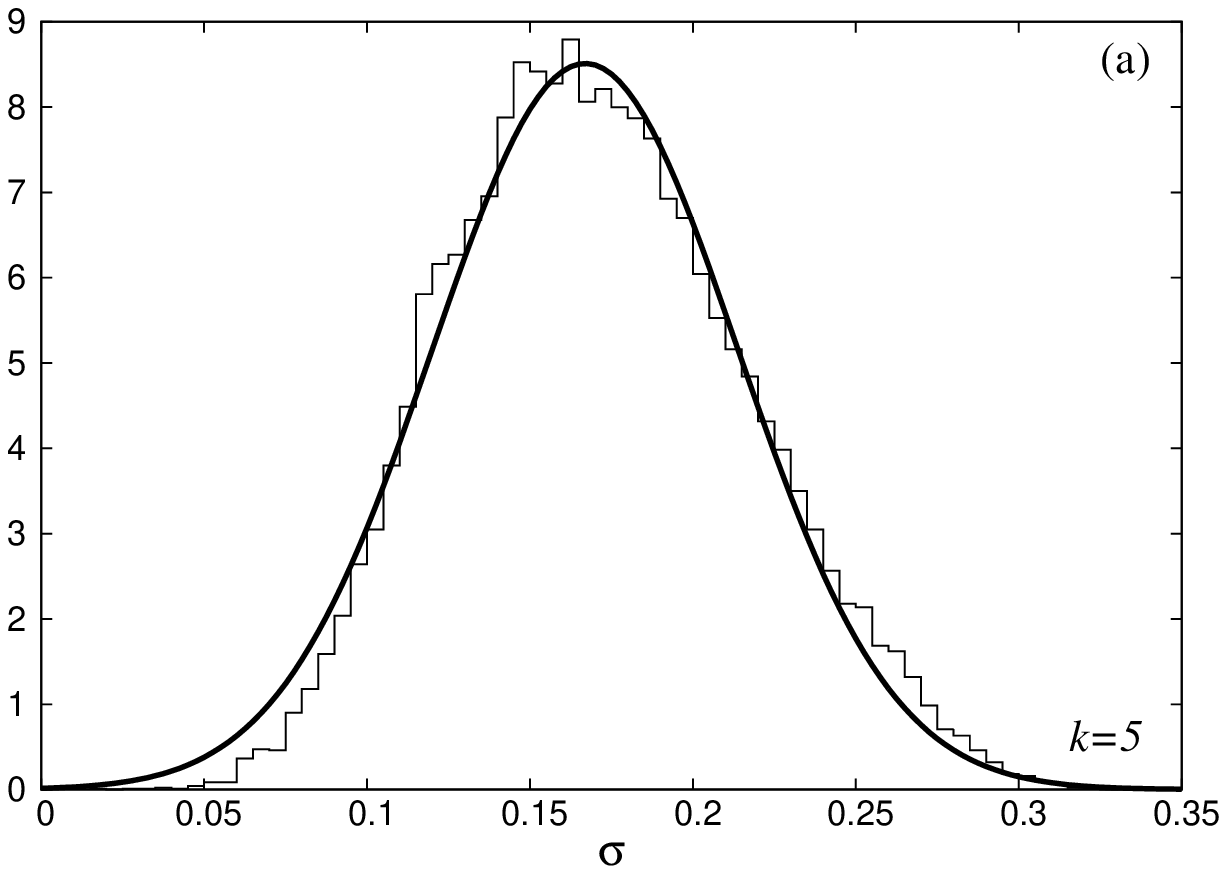}
\includegraphics[width=7.5cm]{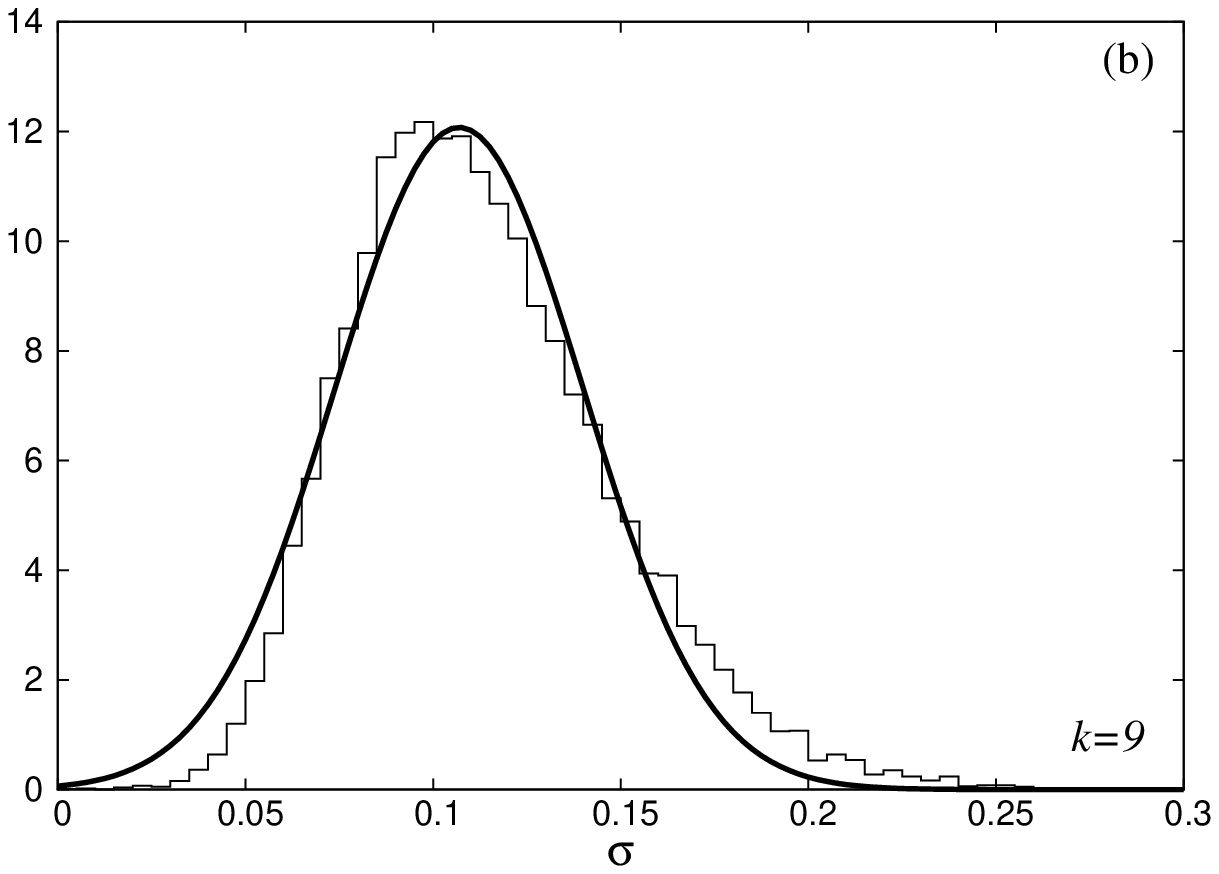}
\includegraphics[width=7.5cm]{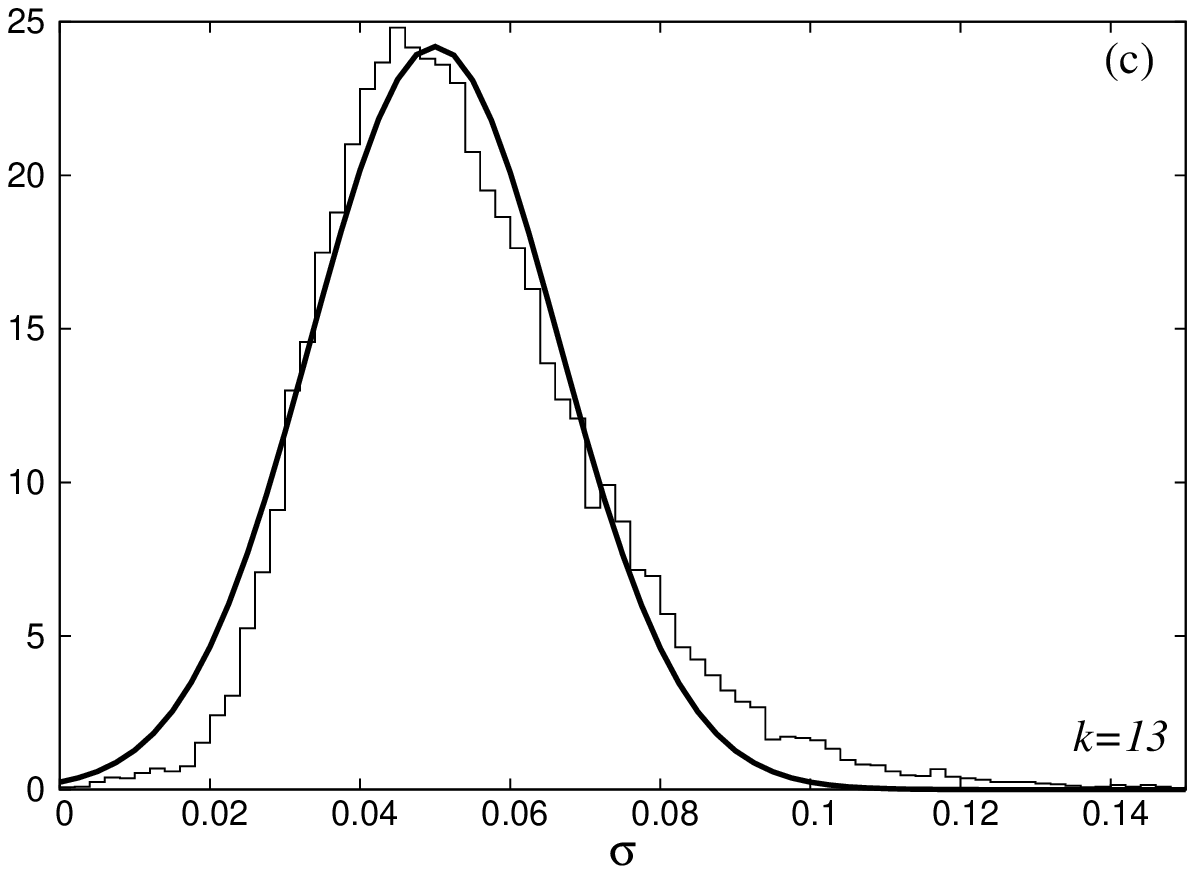}
\includegraphics[width=7.5cm]{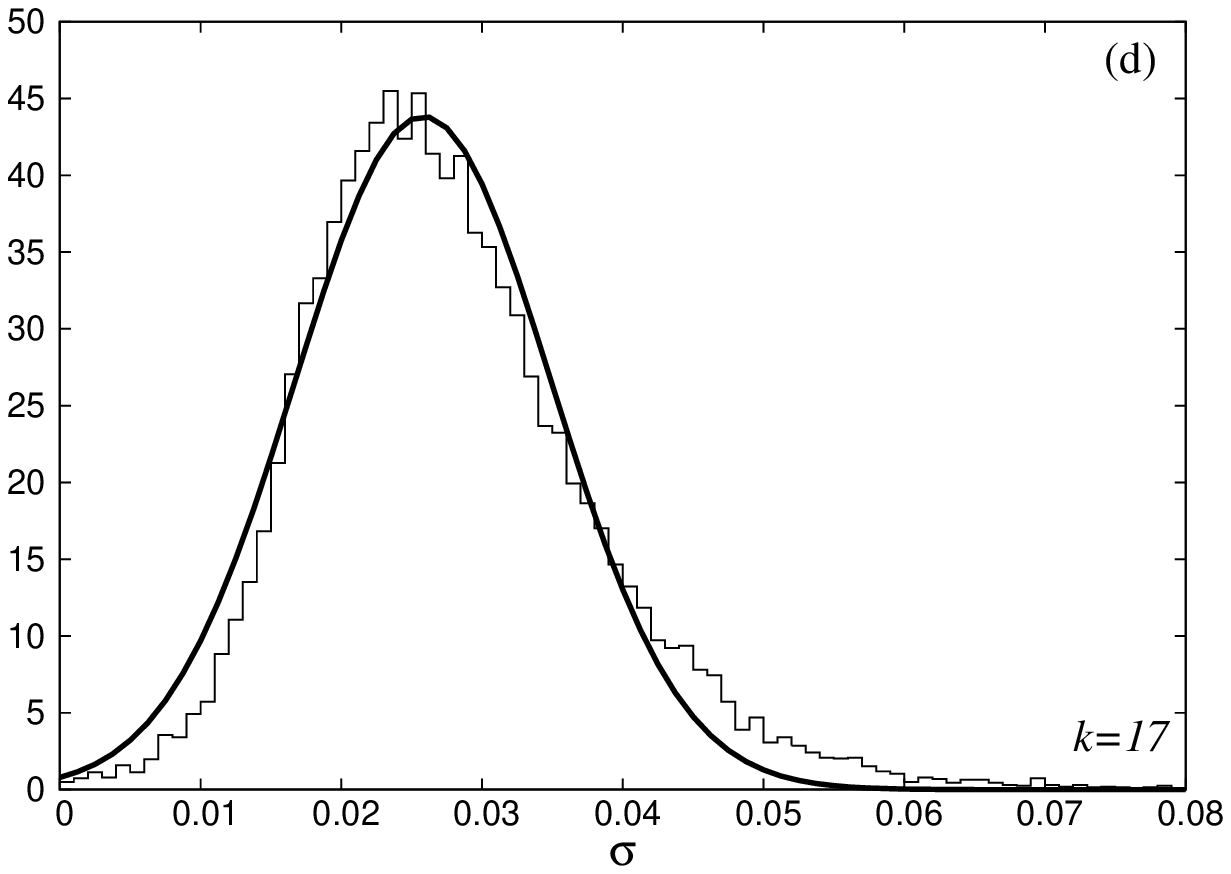}
\caption{We show the histograms of
the slopes $\sigma$ for four systems, matrices of
dimension $N=3000$, for each of them with seven
different values of $k$ close to $k_0=5,9,13,17$,
namely $k=k_0\pm j\delta k$, where $j=0,1,2,3$
and $\delta k=0.00125$:
(a) $k_0=5$, (b) $k_0=9$, (c) $k_0=13$ and (d) $k_0=17$.}
\label{fig8}
\end{figure}

We have thereby demonstrated that the empirical localization measures are well defined, while the theoretical prediction for their mean values is not good enough. The reason is that the localization measures of a given fixed system (with fixed $K=10$ and $k$) have a distribution with nonvanishing variance,
which is out of the scope of current semiclassical theories, as they do not predict this distribution and the corresponding variance. This finding, as the central result of the paper \cite{MR2015}, is demonstrated in Fig.~\ref{fig8}. The distributions are clearly seen to be close to a Gaussian, but cannot be exactly that, as $\sigma$ is always a positive definite quantity. Its inverse, the localization length equal to $\linf=2/\sigma$, has a distribution whose empirical histograms are much further away from a Gaussian, so that in this sense $\sigma$ is the fundamental quantity. Indeed, as we will see, it corresponds to the  finite time Lyapunov exponent known in the theory of dynamical systems.

\begin{figure*}
\center
\includegraphics[width=7.5cm]{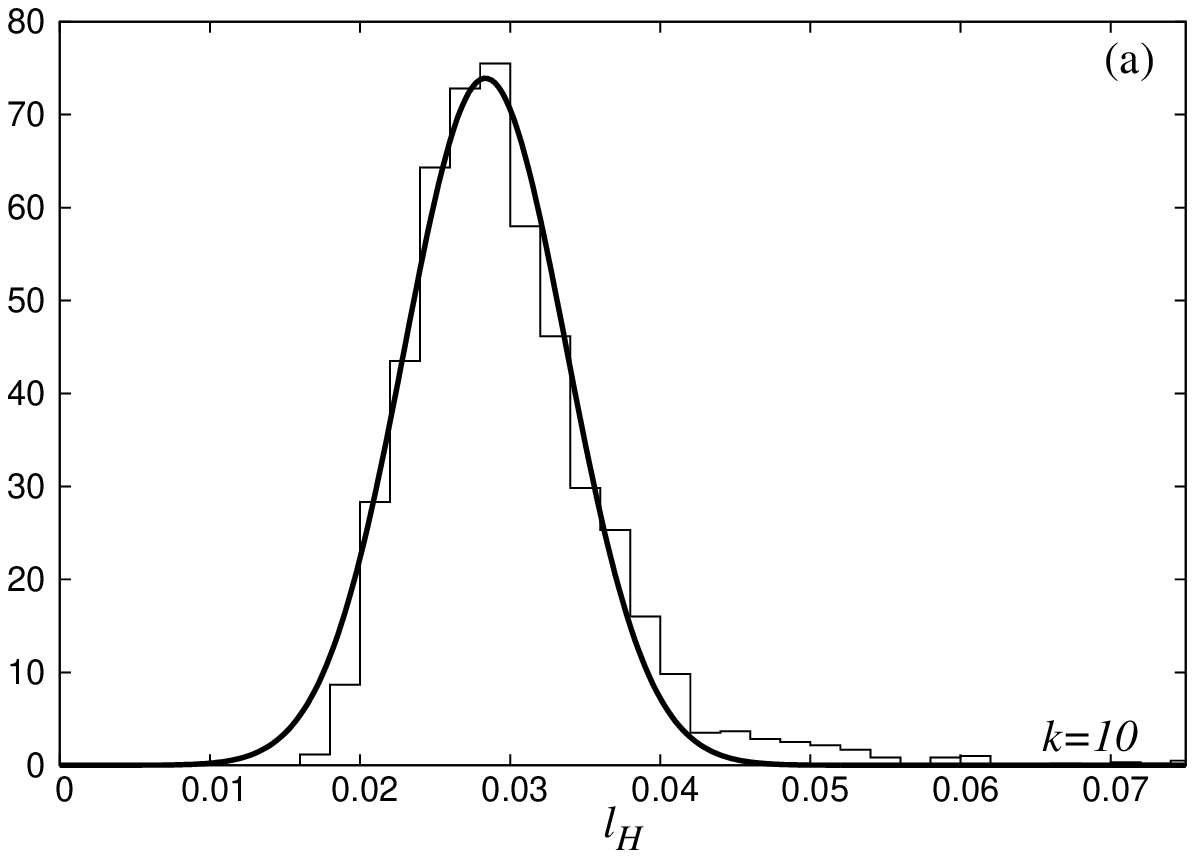}
\includegraphics[width=7.5cm]{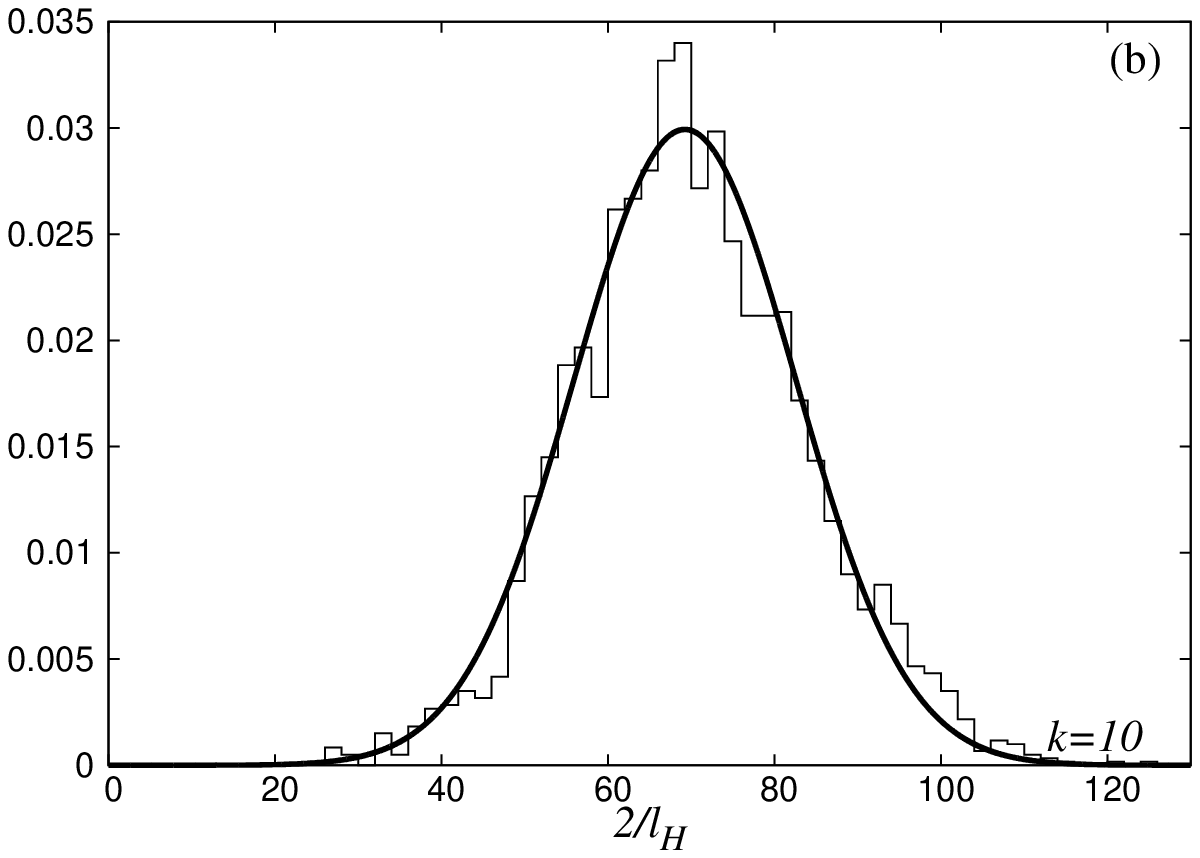}
\caption{We show the histograms of
$l_H$ in (a) and $2/\l_H$ in (b) for the system $k=10$ described by the matrices of dimension $N=3000$. In both cases we show the Gaussian best fit.}
\label{fig9}
\end{figure*}

As $l_H$ and  $2/\sigma$ are equivalent localization measures, the former one is expected also to have a distribution, which we demonstrate in the histograms of Fig.~\ref{fig9}.

\begin{figure*}
\center
\includegraphics[width=7.5cm]{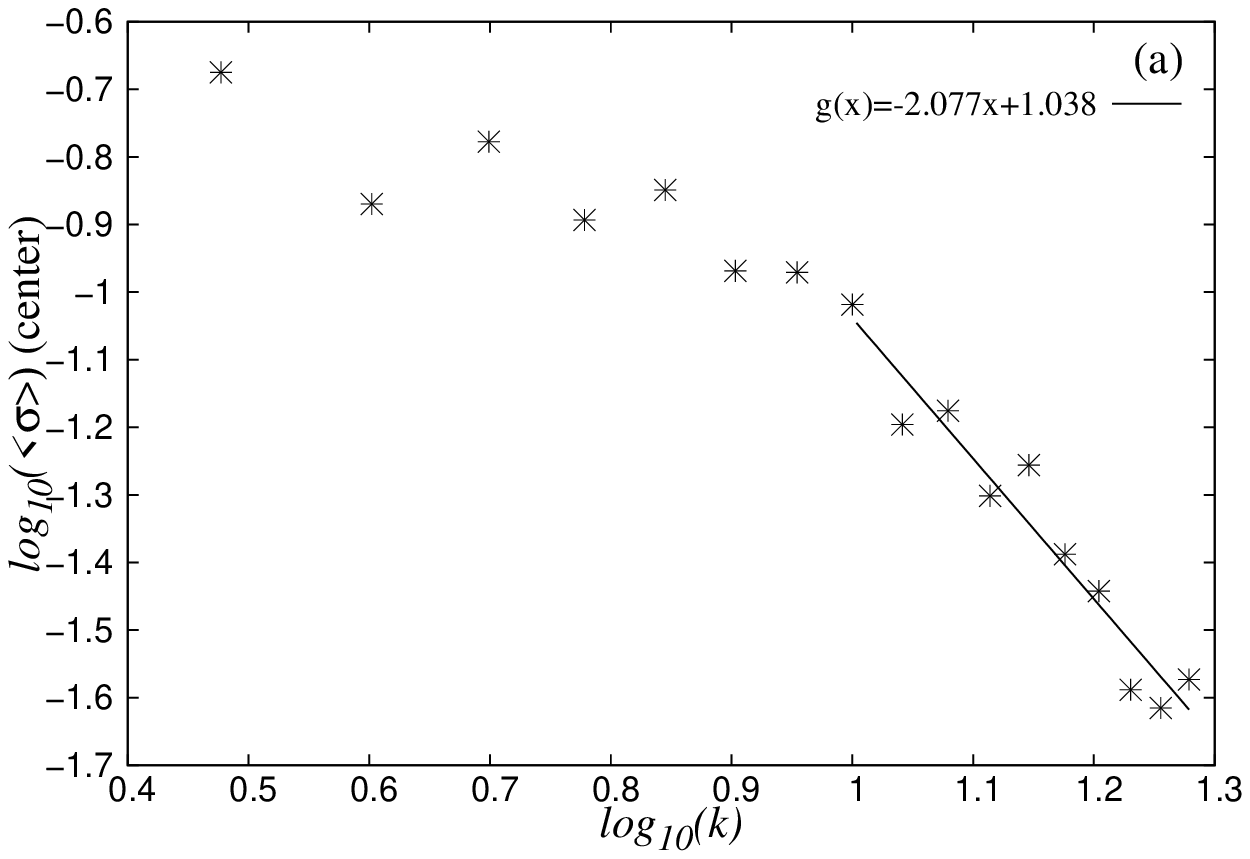}
\includegraphics[width=7.5cm]{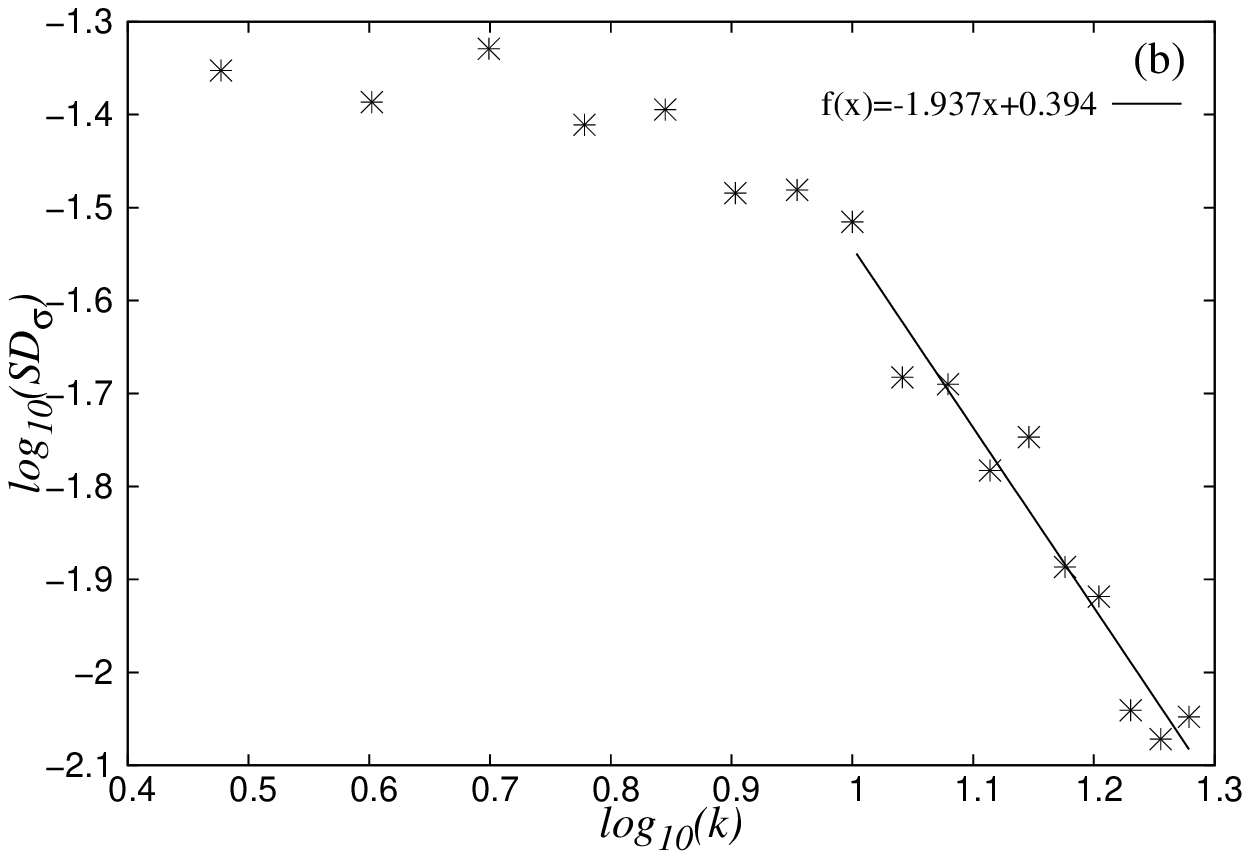}
\caption{We show log-log plots in (a) the mean slope
$\langle \sigma \rangle$ as a function of $k$, and in (b) the standard deviation of $\sigma$ as a function of $k$. The fitting by a straight line is only on the semiclassical interval $10\le k\le 19$. In the former case the behavior is roughly as $1/k^2$, in agreement with the theoretical estimate $1/k^2$ of Eq.~(\ref{finallinf}), and in the latter case also like $1/k^2$, surely not as the theoretical estimate $1/k$ based on the Lyapunov
exponents method in the reference \cite{Kottos1996} [Eq.~(9)].}
\label{fig10}
\end{figure*}

We have also analyzed how the localization measures vary in the semiclassical limit of the increasing value of the quantum parameter $k$, at fixed classical parameter $K=10$. Indeed, the theoretical estimate of $\linf$ in Eq.~(\ref{finallinf}), at fixed $K$, and remembering $k=K/\tau$, shows that approximately the mean value of the localization length should increase quadratically with $k$, or equivalently, the slope $\sigma$ should decrease inversely quadratically with $k$. This prediction is observed, and is demonstrated in the Table~\ref{tabManRob2014-1}, and also in Fig.~\ref{fig10}.

It is also in agreement with the prediction based on the tight-binding approximations in reference \cite{Kottos1996} [Eq.~(6)]. We give, in Table~ \ref{tabManRob2014-1}, the mean slope $\sigma$ and the standard deviation of $\sigma$, as well as the mean value of the related quantity $2/l_H$ and its standard deviation for various $k=k_0=3,4,5,\dots,19$, for each of them taking
seven nearby values of $k$, namely $k=k_0\pm j\delta k$, where $j=0,1,2,3$ and $\delta k =0.00125$, for matrices of dimension $N=3000$. Each histogram for all $k_0$ was fitted with the Gaussian distribution and then the mean values and the standard deviations were extracted. All four quantities decrease to zero with increasing $k$, meaning that in the semiclassical limit the localization
lengths monotonically increase to infinity, so that in this limit we have asymptotically extended states (no localization), and their standard deviation also goes to zero as $1/k^2$, which is different from the tight-binding approximations in reference \cite{Kottos1996} [Eq.~(9)].

\begin{table}\caption{\label{tabManRob2014-1}}
\textbf{The mean value and the standard deviation of the slopes $\sigma$
and $2/l_H$ as a function of $k=k_0=3,4,5,\dots,19$. For each $k=k_0$ we used $N=7\times 3000$ slopes $\sigma$ (see text). All quantities decay to zero in the semiclassical limit.}
\begin{ruledtabular}
\begin{tabular}{cccccc}
\multicolumn{5}{c}{$K=10$ \ -- \ $N=7\times3000$~(slopes) -- \ $N=3000~(2/l_H)$} \\
$k$ & $<\sigma>$ & $SD_{\sigma}$ & $<2/l_H>$ & $SD_{2/l_H}$\\
\colrule
3  & 0.06209 & 0.01324 & 0.062098 & 0.01324 \\
4  & 0.04327 & 0.01073 & 0.043272 & 0.01073 \\
5  & 0.04636 & 0.00758 & 0.046363 & 0.00758 \\
6  & 0.04030 & 0.00974 & 0.040303 & 0.00974 \\
7  & 0.04095 & 0.00838 & 0.040954 & 0.00838 \\
8  & 0.03004 & 0.00756 & 0.030047 & 0.00756 \\
9  & 0.03174 & 0.00600 & 0.031743 & 0.00600 \\
10 & 0.02835 & 0.00539 & 0.028355 & 0.00539 \\
11 & 0.02034 & 0.00353 & 0.020341 & 0.00353 \\
12 & 0.02014 & 0.00321 & 0.020143 & 0.00321 \\
13 & 0.01719 &  0.0029 & 0.017193 & 0.00294 \\
14 & 0.01750 & 0.00289 & 0.017509 & 0.00289 \\
15 & 0.01356 & 0.00230 & 0.013569 & 0.00230 \\
16 & 0.01221 & 0.00194 & 0.012213 & 0.00194 \\
17 & 0.00978 & 0.00148 & 0.009787 & 0.00148 \\
18 & 0.00855 & 0.00128 & 0.008550 & 0.00128 \\
19 & 0.00975 & 0.00141 & 0.009754 & 0.00141 \\
\end{tabular}
\end{ruledtabular}
\end{table}

Next we want to study how does the distribution of the localization measure
$\sigma$ behave as a function of the dimension $N$ of the Izrailev model
Eqs.~(\ref{u_repres}-\ref{Bnmoper}). Since in the limit $N\rightarrow \infty$ the model converges to the infinitely dimensional quantum kicked rotator, we would at first sight expect that following the Shepelyansky picture \cite{She1986} $\sigma$ should converge to its asymptotic value, which is sharply defined in the sense that the variance of the distribution of $\sigma$ goes to zero inversely with $N$. Namely, at fixed $K$ and $k$ Shepelyansky reduces the problem of calculating the localization length to the problem of the finite time Lyapunov exponents of the {\bf approximate} underlying finite dimensional Hamilton system with dimension $2k$. The localization length is then found to be equal to the inverse value of the smallest positive Lyapunov exponent.  In our case, the dimension of the matrices $N$ of the Izrailev model plays the role of time.  As it is known, and analyzed in detail in the paper
\cite{MR2015}, the finite time Lyapunov exponents have a distribution, which is almost Gaussian, and its variance decays to zero inversely with time. Thus on the basis of this we would expect that the variance of $\sigma$ decays inversely with $N$.

However, this is not what we observe. In the Table~\ref{tabManRob2014-2}
we clearly see that at constant $K=10$ and $k=10$ the mean value of $\sigma$
is constant and obviously equal to its asymptotic value of $N=\infty$,
while the variance of $\sigma$ does not decrease with $N$, as $1/N$,
but is constant instead, independent of $N$. This is in disagreement with
the banded-matrix models of the tight-binding approximations and thus
disagrees with the Eq.~(9) of reference \cite{Kottos1996}, and also disagrees with the Shepelyansky picture. The reason  is that the associated Shepelyansky's Hamilton system is only approximate construction, because with increasing $N$ the matrix elements of the Floquet propagator (matrix) outside the diagonal band of width $2k$ become important, and thus the dimension of
the Hamilton system cannot be considered finite, constant and equal to $2k$,
but increases with $N$. As a consequence we have the constant value of the variance of $\sigma$, and thus constant variance of the localization length ${\cal L}=2/\sigma$, and therefore the localization length has a distribution with nonvanishing variance even in the limit $N=\infty$. This is precisely the reason why the semiclassical prediction of the localization length in
Eq.~(\ref{finallinf}) fails in detail and we find strong fluctuations in the plot of $\linf$ against the $2/\sigma$ of Figs.~\ref{fig4} and \ref{fig7}. The proper theory of the localization length must predict its distribution rather than just its approximate mean value.

\begin{table}\caption{\label{tabManRob2014-2}}
\textbf{The mean value and the variance of the slope $\sigma$ as
a function of the matrix dimension $N$ for a fixed system
with $K=10$ and $k=10$. Both are obviously constant.}
\begin{ruledtabular}
\begin{tabular}{cccccc}
\multicolumn{3}{c}{$K=10$ \ -- $k=10$ } \\
$N$ & $\langle \sigma \rangle$ & $var_{\sigma}$ \\
\colrule
500   & 0.102624 & 0.00113224 \\
1000  & 0.101170 & 0.00112558 \\
2000  & 0.100066 & 0.00115575 \\
3000  & 0.102217 & 0.00110438 \\
\end{tabular}
\end{ruledtabular}
\end{table}

As it is well known the problem of quantum or dynamical localization is related to the Anderson localization model, within the framework of the tight-binding approximation, with hopping transitions between the nearest neighbors only. This goes back to the pioneering work of Fishman \textit{et al.} \cite{FGP1982}, as discussed in \cite{Haake,Stoe}, and also reviewed in \cite{PGF1984}.

\begin{figure}
\center
\includegraphics[width=7.5cm]{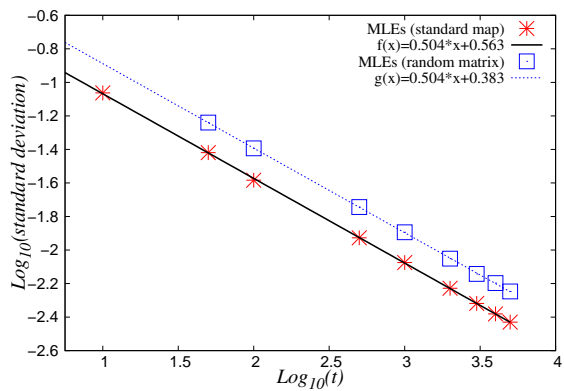}
\caption{[Color online] The standard deviation of the positive finite time
Lyapunov exponents for the  standard map (stars) and  for the product of random transfer matrices with the box distribution of $W$ (empty boxes), as a function of time in $\log-\log$ presentation, and their best fits. The slope is exactly -1/2.}
\label{fig11}
\end{figure}

In the paper \cite{MR2015} we have numerically analyzed the behavior of the
finite time Lyapunov exponents for a classical Hamilton system exemplified by
the standard map (SM), following Fujisaka \cite{Fuj1983} and Ott \cite{Ott1993} and also for the $2\times 2$ random transfer matrices of the tight-binding approximation to describe the Anderson localization. In both cases we have shown that the distribution of the positive Lyapunov exponent is excellently described by a Gaussian distribution, whose mean value converges with time to the asymptotic value of the infinite time, while the variance decays inversely
with time $t$ (the number of iterations in the case of the standard map),
and $n$, the number of random matrices in the product. The latter are random unimodular transfer matrices of the tight-binding approximation, of the form
\be \label{transfermat}
T_n =
\left(
\ba {cc}
W & -1 \\
1 & 0
\ea
\right)
\ee
where $W=E-E_n^0$ is drawn from a distribution, defined by a given model. $E$ is the eigenenergy of the system, and $E_n^0$ is the fluctuating on-site potential. We have tested three quite different distributions for $W$, namely Gaussian, box distribution and the Cauchy-Lorentz distribution \cite{MR2015}, and found that the shape of the distribution of the positive Lyapunov exponent for any $n$ (=number of matrices in the product) depends very weakly on the overall shape of the $W$-distribution, while the mean value and the variance depend only on
the variance of $W$. Indeed, the evidence for the predicted decay of the variance of the finite time Lyapunov exponents is overwhelming, as shown in Fig.~\ref{fig11}, where we plot the standard deviation as a function of time in log-log plot, showing that it decays inversely with the square root of time.

In the context of our Izrailev model the dimension $N$ of the matrix plays the role of time. The width of the diagonal band is equal to $2k$. Shepelyansky reduces the problem of the localization length to the determination of the smallest positive Lyapunov exponent (its inverse is the localization length)
of the underlying finite dimensional Hamilton system with dimension $2k$. Then, the finite time Lyapunov exponent should have some almost Gaussian distribution, whose mean tends to the asymptotic Lyapunov exponent with $N\rightarrow \infty$ and the variance should decrease to zero as $1/N$.

If this picture were exact, then the mean localization length as a function of $N$ should converge to the asymptotic value, which we do observe in Table~ \ref{tabManRob2014-2} of Sec.~\ref{sec3}, while the variance does {\em not}
decay to zero, but rather remains constant, independent of $N$ as clearly demonstrated in Table \ref{tabManRob2014-2}. From this we conclude that even in the limit $N\rightarrow \infty$ the localization length has a certain distribution with nonvanishing variance, or more precisely, its inverse (the slope $\sigma$) has an almost Gaussian distribution with nonvanishing variance. We believe that this is the cause of the strong fluctuations observed for example in Fig.~\ref{fig4} and \ref{fig7}(a) of Sec.~\ref{sec2}.

\section{Summary} \label{sec4}

We have reviewed our recent results \cite{MR2013,MR2014,MR2015}
on the dynamical localization in the $N$-dimensional Izrailev model. The analysis of the classical system (standard map) and of the quantum kicked rotator (QKR) has been performed  for many different values of the classical
kick parameter $K$ on the interval $[K_{cr},70]\approx [0.97,70]$, of the quantum parameter $k$ and matrix dimensions $N$ in the interval $[400,3000]$. The aspects of classical generally anomalous diffusion have been studied and the important relevance of the accelerator modes elucidated, and the
semiclassical approximation for the average localization length $\linf$ has been derived. The entropy localization measure $l_H$ has been calculated,
and the corresponding parameter $\blo= \langle l_H\rangle/N$ defined.
The scaling law of $\blo$ versus $\La=\linf/N$ was found, in agreement
with the previous results of Izrailev (\cite{Izr1990} and references therein).
However, even after a great numerical effort in maximally improving the
statistical relevance of this scaling law, large fluctuations around the
mean value have been observed. Also, we have shown, that the Brody distribution describes the level spacing distribution very well, and the spectral Brody parameter $\bro$, determining the level spacing distribution, was found to be linearly related to $\blo$.

In the end our main conclusion is the empirical fact based on our numerical computations of the eigenfunctions of the $N$-dimensional Izrailev model, that the localization length has a distribution with nonvanishing variance not
only for finite $N$, but even in the limit $N\rightarrow \infty$. This is the reason, we believe, for the strong fluctuations in the scaling laws which involve the empirical localization measures and the theoretical semiclassical value of the localization length. In the Shepelyansky picture \cite{She1986}
this might seem to be a contradiction, but the resolution of the puzzle is that in the limit of large $N$ the finite dimensional Hamilton system extracted from the Floquet propagator of the quantum kicked rotator is not good enough, and therefore the matrix elements outside the main diagonal band of width $2k$ play a role, even if they are small, but nevertheless plentiful, making the Hamilton system effectively infinite dimensional, with infinitely many Lyapunov exponents. This finding is a challenge for the improved semiclassical theory of the localization length, to derive and explain the discovered distribution function. On the other hand, the simple model of the Anderson localization based on the tight-binding approximation, with only the nearest neighbor interactions,  described by the product of $2\times2$ unimodular matrices, has a finite dimension, as the transfer matrices are exactly two-dimensional, and therefore the variance vanishes in the limit of large times $n$ (number of matrices in the product)  as $1/n$. The same conclusion applies to such a model with a finite number of interacting neighbors. Indeed, according to the references \cite{Kottos1996,Kottos1999} the variance of $\sigma$ should vanish as $Var(\sigma) \propto 1/(Nk^2)$, but our work shows that in the quantum kicked rotator this is not observed: the variance does not depend on $N$, and decays with $k$ faster than $1/k^2$, namely as $1/k^4$. Thus, we found some important differences between the dynamical localization in the quantum kicked rotator and the Anderson tight-binding model of localization, and the Shepelyansky picture, which rest upon the banded matrix models with rigorously finite bandwidth. Therefore, the problem of calculating the distribution of the localization length (or its inverse) in the semiclassical framework is open for the future work. Also, the derivation of the Brody distribution to explain the level spacing distribution of the energies \cite{BatRob2010,BatRob2013,BatRob2013A} in time-independent systems, and of
the quasienergies \cite{Izr1990,MR2013,BMR2013} in time-periodic systems,
of chaotic eigenstates, is still open for the future.

\section*{Acknowledgements}
This work was supported by the Slovenian Research Agency (ARRS)
under the grant J1-4004.

\bibliography{lj2014FR}

\providecommand{\noopsort}[1]{}\providecommand{\singleletter}[1]{#1}%
\begin{thebibliography}{65}%
\makeatletter
\providecommand \@ifxundefined [1]{%
 \@ifx{#1\undefined}
}%
\providecommand \@ifnum [1]{%
 \ifnum #1\expandafter \@firstoftwo
 \else \expandafter \@secondoftwo
 \fi
}%
\providecommand \@ifx [1]{%
 \ifx #1\expandafter \@firstoftwo
 \else \expandafter \@secondoftwo
 \fi
}%
\providecommand \natexlab [1]{#1}%
\providecommand \enquote  [1]{``#1''}%
\providecommand \bibnamefont  [1]{#1}%
\providecommand \bibfnamefont [1]{#1}%
\providecommand \citenamefont [1]{#1}%
\providecommand \href@noop [0]{\@secondoftwo}%
\providecommand \href [0]{\begingroup \@sanitize@url \@href}%
\providecommand \@href[1]{\@@startlink{#1}\@@href}%
\providecommand \@@href[1]{\endgroup#1\@@endlink}%
\providecommand \@sanitize@url [0]{\catcode `\\12\catcode `\$12\catcode
  `\&12\catcode `\#12\catcode `\^12\catcode `\_12\catcode `\%12\relax}%
\providecommand \@@startlink[1]{}%
\providecommand \@@endlink[0]{}%
\providecommand \url  [0]{\begingroup\@sanitize@url \@url }%
\providecommand \@url [1]{\endgroup\@href {#1}{\urlprefix }}%
\providecommand \urlprefix  [0]{URL }%
\providecommand \Eprint [0]{\href }%
\providecommand \doibase [0]{http://dx.doi.org/}%
\providecommand \selectlanguage [0]{\@gobble}%
\providecommand \bibinfo  [0]{\@secondoftwo}%
\providecommand \bibfield  [0]{\@secondoftwo}%
\providecommand \translation [1]{[#1]}%
\providecommand \BibitemOpen [0]{}%
\providecommand \bibitemStop [0]{}%
\providecommand \bibitemNoStop [0]{.\EOS\space}%
\providecommand \EOS [0]{\spacefactor3000\relax}%
\providecommand \BibitemShut  [1]{\csname bibitem#1\endcsname}%
\let\auto@bib@innerbib\@empty
\bibitem [{\citenamefont {St\"ockmann}(1999)}]{Stoe}%
  \BibitemOpen
  \bibfield  {author} {\bibinfo {author} {\bibfnamefont {H.~J.}\ \bibnamefont
  {St\"ockmann}},\ }\href@noop {} {\emph {\bibinfo {title} {Quantum Chaos - An
  Introduction}}}\ (\bibinfo  {publisher} {Cambridge: Cambridge University
  Press},\ \bibinfo {year} {1999})\BibitemShut {NoStop}%
\bibitem [{\citenamefont {Haake}(2001)}]{Haake}%
  \BibitemOpen
  \bibfield  {author} {\bibinfo {author} {\bibfnamefont {F.}~\bibnamefont
  {Haake}},\ }\href@noop {} {\emph {\bibinfo {title} {Quantum Signatures of
  Chaos}}}\ (\bibinfo  {publisher} {Berlin: Springer},\ \bibinfo {year}
  {2001})\BibitemShut {NoStop}%
\bibitem [{\citenamefont {Casati}\ \emph {et~al.}(1979)\citenamefont {Casati},
  \citenamefont {Chirikov}, \citenamefont {Ford},\ and\ \citenamefont
  {Izrailev}}]{CCFI79}%
  \BibitemOpen
  \bibfield  {author} {\bibinfo {author} {\bibfnamefont {G.}~\bibnamefont
  {Casati}}, \bibinfo {author} {\bibfnamefont {B.}~\bibnamefont {Chirikov}},
  \bibinfo {author} {\bibfnamefont {J.}~\bibnamefont {Ford}}, \ and\ \bibinfo
  {author} {\bibfnamefont {F.~M.}\ \bibnamefont {Izrailev}},\ }\href@noop {}
  {\bibfield  {journal} {\bibinfo  {journal} {Lect. Notes Phys.}\ }\textbf
  {\bibinfo {volume} {93}},\ \bibinfo {pages} {334} (\bibinfo {year}
  {1979})}\BibitemShut {NoStop}%
\bibitem [{\citenamefont {Robnik}(1998)}]{Rob1998}%
  \BibitemOpen
  \bibfield  {author} {\bibinfo {author} {\bibfnamefont {M.}~\bibnamefont
  {Robnik}},\ }\href@noop {} {\bibfield  {journal} {\bibinfo  {journal} {Nonl.
  Phen. in Compl. Syst. (Minsk)}\ }\textbf {\bibinfo {volume} {1}},\ \bibinfo
  {pages} {1} (\bibinfo {year} {1998})}\BibitemShut {NoStop}%
\bibitem [{\citenamefont {Mehta}(1991)}]{Mehta}%
  \BibitemOpen
  \bibfield  {author} {\bibinfo {author} {\bibfnamefont {M.~L.}\ \bibnamefont
  {Mehta}},\ }\href@noop {} {\emph {\bibinfo {title} {Random Matrices}}}\
  (\bibinfo  {publisher} {Boston: Academic Press},\ \bibinfo {year}
  {1991})\BibitemShut {NoStop}%
\bibitem [{\citenamefont {Guhr}\ \emph {et~al.}(1998)\citenamefont {Guhr},
  \citenamefont {M\"uller-Groeling},\ and\ \citenamefont
  {Weidenm\"uller}}]{GMW}%
  \BibitemOpen
  \bibfield  {author} {\bibinfo {author} {\bibfnamefont {T.}~\bibnamefont
  {Guhr}}, \bibinfo {author} {\bibfnamefont {A.}~\bibnamefont
  {M\"uller-Groeling}}, \ and\ \bibinfo {author} {\bibfnamefont
  {H.}~\bibnamefont {Weidenm\"uller}},\ }\href@noop {} {\bibfield  {journal}
  {\bibinfo  {journal} {Phys. Rep.}\ }\textbf {\bibinfo {volume} {299}},\
  \bibinfo {pages} {4} (\bibinfo {year} {1998})}\BibitemShut {NoStop}%
\bibitem [{\citenamefont {Bohigas}\ \emph {et~al.}(1984)\citenamefont
  {Bohigas}, \citenamefont {Giannoni},\ and\ \citenamefont {Schmit}}]{BGS}%
  \BibitemOpen
  \bibfield  {author} {\bibinfo {author} {\bibfnamefont {O.}~\bibnamefont
  {Bohigas}}, \bibinfo {author} {\bibfnamefont {M.~J.}\ \bibnamefont
  {Giannoni}}, \ and\ \bibinfo {author} {\bibfnamefont {C.}~\bibnamefont
  {Schmit}},\ }\href@noop {} {\bibfield  {journal} {\bibinfo  {journal} {Phys.
  Rev. Lett.}\ }\textbf {\bibinfo {volume} {52}},\ \bibinfo {pages} {1}
  (\bibinfo {year} {1984})}\BibitemShut {NoStop}%
\bibitem [{\citenamefont {Casati}\ \emph {et~al.}(1980)\citenamefont {Casati},
  \citenamefont {Valz-Gris},\ and\ \citenamefont {Guarneri}}]{Cas}%
  \BibitemOpen
  \bibfield  {author} {\bibinfo {author} {\bibfnamefont {G.}~\bibnamefont
  {Casati}}, \bibinfo {author} {\bibfnamefont {F.}~\bibnamefont {Valz-Gris}}, \
  and\ \bibinfo {author} {\bibfnamefont {I.}~\bibnamefont {Guarneri}},\
  }\href@noop {} {\bibfield  {journal} {\bibinfo  {journal} {Lett. Nuovo
  Cimento}\ }\textbf {\bibinfo {volume} {28}},\ \bibinfo {pages} {279}
  (\bibinfo {year} {1980})}\BibitemShut {NoStop}%
\bibitem [{\citenamefont {Robnik}\ and\ \citenamefont {Berry}(1986)}]{RB1986}%
  \BibitemOpen
  \bibfield  {author} {\bibinfo {author} {\bibfnamefont {M.}~\bibnamefont
  {Robnik}}\ and\ \bibinfo {author} {\bibfnamefont {M.~V.}\ \bibnamefont
  {Berry}},\ }\href@noop {} {\bibfield  {journal} {\bibinfo  {journal} {J.
  Phys. A: Math. Gen.}\ }\textbf {\bibinfo {volume} {19}},\ \bibinfo {pages}
  {669} (\bibinfo {year} {1986})}\BibitemShut {NoStop}%
\bibitem [{\citenamefont {Robnik}(1986)}]{Rob1986}%
  \BibitemOpen
  \bibfield  {author} {\bibinfo {author} {\bibfnamefont {M.}~\bibnamefont
  {Robnik}},\ }\href@noop {} {\bibfield  {journal} {\bibinfo  {journal} {Lect.
  Notes Phys.}\ }\textbf {\bibinfo {volume} {263}},\ \bibinfo {pages} {120}
  (\bibinfo {year} {1986})}\BibitemShut {NoStop}%
\bibitem [{\citenamefont {Berry}(1985)}]{Berry1985}%
  \BibitemOpen
  \bibfield  {author} {\bibinfo {author} {\bibfnamefont {M.~V.}\ \bibnamefont
  {Berry}},\ }\href@noop {} {\bibfield  {journal} {\bibinfo  {journal} {Proc.
  Roy. Soc. Lond. A}\ }\textbf {\bibinfo {volume} {400}},\ \bibinfo {pages}
  {229} (\bibinfo {year} {1985})}\BibitemShut {NoStop}%
\bibitem [{\citenamefont {Sieber}\ and\ \citenamefont
  {Richter}(2001)}]{Sieber}%
  \BibitemOpen
  \bibfield  {author} {\bibinfo {author} {\bibfnamefont {M.}~\bibnamefont
  {Sieber}}\ and\ \bibinfo {author} {\bibfnamefont {K.}~\bibnamefont
  {Richter}},\ }\href@noop {} {\bibfield  {journal} {\bibinfo  {journal} {Phys.
  Scr.}\ }\textbf {\bibinfo {volume} {T90}},\ \bibinfo {pages} {128} (\bibinfo
  {year} {2001})}\BibitemShut {NoStop}%
\bibitem [{\citenamefont {M\"uller}\ \emph {et~al.}(2004)\citenamefont
  {M\"uller}, \citenamefont {Heusler}, \citenamefont {Braun}, \citenamefont
  {Haake},\ and\ \citenamefont {Altland}}]{Mueller1}%
  \BibitemOpen
  \bibfield  {author} {\bibinfo {author} {\bibfnamefont {S.}~\bibnamefont
  {M\"uller}}, \bibinfo {author} {\bibfnamefont {S.}~\bibnamefont {Heusler}},
  \bibinfo {author} {\bibfnamefont {P.}~\bibnamefont {Braun}}, \bibinfo
  {author} {\bibfnamefont {F.}~\bibnamefont {Haake}}, \ and\ \bibinfo {author}
  {\bibfnamefont {A.}~\bibnamefont {Altland}},\ }\href@noop {} {\bibfield
  {journal} {\bibinfo  {journal} {Phys. Rev. Lett.}\ }\textbf {\bibinfo
  {volume} {93}},\ \bibinfo {pages} {014103} (\bibinfo {year}
  {2004})}\BibitemShut {NoStop}%
\bibitem [{\citenamefont {Heusler}\ \emph {et~al.}(2004)\citenamefont
  {Heusler}, \citenamefont {M\"uller}, \citenamefont {Braun},\ and\
  \citenamefont {Haake}}]{Mueller2}%
  \BibitemOpen
  \bibfield  {author} {\bibinfo {author} {\bibfnamefont {S.}~\bibnamefont
  {Heusler}}, \bibinfo {author} {\bibfnamefont {S.}~\bibnamefont {M\"uller}},
  \bibinfo {author} {\bibfnamefont {P.}~\bibnamefont {Braun}}, \ and\ \bibinfo
  {author} {\bibfnamefont {F.}~\bibnamefont {Haake}},\ }\href@noop {}
  {\bibfield  {journal} {\bibinfo  {journal} {J. Phys.A: Math. Gen.}\ }\textbf
  {\bibinfo {volume} {37}},\ \bibinfo {pages} {L31} (\bibinfo {year}
  {2004})}\BibitemShut {NoStop}%
\bibitem [{\citenamefont {M\"uller}\ \emph {et~al.}(2005)\citenamefont
  {M\"uller}, \citenamefont {Heusler}, \citenamefont {Braun}, \citenamefont
  {Haake},\ and\ \citenamefont {Altland}}]{Mueller3}%
  \BibitemOpen
  \bibfield  {author} {\bibinfo {author} {\bibfnamefont {S.}~\bibnamefont
  {M\"uller}}, \bibinfo {author} {\bibfnamefont {S.}~\bibnamefont {Heusler}},
  \bibinfo {author} {\bibfnamefont {P.}~\bibnamefont {Braun}}, \bibinfo
  {author} {\bibfnamefont {F.}~\bibnamefont {Haake}}, \ and\ \bibinfo {author}
  {\bibfnamefont {A.}~\bibnamefont {Altland}},\ }\href@noop {} {\bibfield
  {journal} {\bibinfo  {journal} {Phys. Rev. E}\ }\textbf {\bibinfo {volume}
  {72}},\ \bibinfo {pages} {046207} (\bibinfo {year} {2005})}\BibitemShut
  {NoStop}%
\bibitem [{\citenamefont {M\"uller}\ \emph {et~al.}(2009)\citenamefont
  {M\"uller}, \citenamefont {Heusler}, \citenamefont {Altland}, \citenamefont
  {Braun},\ and\ \citenamefont {Haake}}]{Mueller4}%
  \BibitemOpen
  \bibfield  {author} {\bibinfo {author} {\bibfnamefont {S.}~\bibnamefont
  {M\"uller}}, \bibinfo {author} {\bibfnamefont {S.}~\bibnamefont {Heusler}},
  \bibinfo {author} {\bibfnamefont {A.}~\bibnamefont {Altland}}, \bibinfo
  {author} {\bibfnamefont {P.}~\bibnamefont {Braun}}, \ and\ \bibinfo {author}
  {\bibfnamefont {F.}~\bibnamefont {Haake}},\ }\href@noop {} {\bibfield
  {journal} {\bibinfo  {journal} {New J. of Phys.}\ }\textbf {\bibinfo {volume}
  {11}},\ \bibinfo {pages} {103025} (\bibinfo {year} {2009})}\BibitemShut
  {NoStop}%
\bibitem [{\citenamefont {Robnik}\ and\ \citenamefont {Veble}(1998)}]{RobVeb}%
  \BibitemOpen
  \bibfield  {author} {\bibinfo {author} {\bibfnamefont {M.}~\bibnamefont
  {Robnik}}\ and\ \bibinfo {author} {\bibfnamefont {G.}~\bibnamefont {Veble}},\
  }\href@noop {} {\bibfield  {journal} {\bibinfo  {journal} {J. Phys. A: Math.
  Theor.}\ }\textbf {\bibinfo {volume} {31}},\ \bibinfo {pages} {4669}
  (\bibinfo {year} {1998})}\BibitemShut {NoStop}%
\bibitem [{\citenamefont {Berry}(1977)}]{Berry1977}%
  \BibitemOpen
  \bibfield  {author} {\bibinfo {author} {\bibfnamefont {M.~V.}\ \bibnamefont
  {Berry}},\ }\href@noop {} {\bibfield  {journal} {\bibinfo  {journal} {J.
  Phys. A: Math. Gen.}\ }\textbf {\bibinfo {volume} {12}},\ \bibinfo {pages}
  {2083} (\bibinfo {year} {1977})}\BibitemShut {NoStop}%
\bibitem [{\citenamefont {Berry}\ and\ \citenamefont {Robnik}(1984)}]{BR1984}%
  \BibitemOpen
  \bibfield  {author} {\bibinfo {author} {\bibfnamefont {M.~V.}\ \bibnamefont
  {Berry}}\ and\ \bibinfo {author} {\bibfnamefont {M.}~\bibnamefont {Robnik}},\
  }\href@noop {} {\bibfield  {journal} {\bibinfo  {journal} {J. Phys. A: Math.
  Gen.}\ }\textbf {\bibinfo {volume} {17}},\ \bibinfo {pages} {2413} (\bibinfo
  {year} {1984})}\BibitemShut {NoStop}%
\bibitem [{\citenamefont {Prosen}\ and\ \citenamefont
  {Robnik}(1999)}]{ProRob1999}%
  \BibitemOpen
  \bibfield  {author} {\bibinfo {author} {\bibfnamefont {T.}~\bibnamefont
  {Prosen}}\ and\ \bibinfo {author} {\bibfnamefont {M.}~\bibnamefont
  {Robnik}},\ }\href@noop {} {\bibfield  {journal} {\bibinfo  {journal} {J.
  Phys. A: Math. Gen.}\ }\textbf {\bibinfo {volume} {32}},\ \bibinfo {pages}
  {1863} (\bibinfo {year} {1999})}\BibitemShut {NoStop}%
\bibitem [{\citenamefont {Prosen}\ and\ \citenamefont
  {Robnik}(1993{\natexlab{a}})}]{ProRob1993a}%
  \BibitemOpen
  \bibfield  {author} {\bibinfo {author} {\bibfnamefont {T.}~\bibnamefont
  {Prosen}}\ and\ \bibinfo {author} {\bibfnamefont {M.}~\bibnamefont
  {Robnik}},\ }\href@noop {} {\bibfield  {journal} {\bibinfo  {journal} {J.
  Phys. A: Math. Gen.}\ }\textbf {\bibinfo {volume} {26}},\ \bibinfo {pages}
  {2371} (\bibinfo {year} {1993}{\natexlab{a}})}\BibitemShut {NoStop}%
\bibitem [{\citenamefont {Prosen}\ and\ \citenamefont
  {Robnik}(1993{\natexlab{b}})}]{ProRob1993b}%
  \BibitemOpen
  \bibfield  {author} {\bibinfo {author} {\bibfnamefont {T.}~\bibnamefont
  {Prosen}}\ and\ \bibinfo {author} {\bibfnamefont {M.}~\bibnamefont
  {Robnik}},\ }\href@noop {} {\bibfield  {journal} {\bibinfo  {journal} {J.
  Phys. A: Math. Gen.}\ }\textbf {\bibinfo {volume} {26}},\ \bibinfo {pages}
  {1105} (\bibinfo {year} {1993}{\natexlab{b}})}\BibitemShut {NoStop}%
\bibitem [{\citenamefont {Prosen}\ and\ \citenamefont
  {Robnik}(1994{\natexlab{a}})}]{ProRob1994a}%
  \BibitemOpen
  \bibfield  {author} {\bibinfo {author} {\bibfnamefont {T.}~\bibnamefont
  {Prosen}}\ and\ \bibinfo {author} {\bibfnamefont {M.}~\bibnamefont
  {Robnik}},\ }\href@noop {} {\bibfield  {journal} {\bibinfo  {journal} {J.
  Phys. A: Math. Gen.}\ }\textbf {\bibinfo {volume} {27}},\ \bibinfo {pages}
  {L459} (\bibinfo {year} {1994}{\natexlab{a}})}\BibitemShut {NoStop}%
\bibitem [{\citenamefont {Prosen}\ and\ \citenamefont
  {Robnik}(1994{\natexlab{b}})}]{ProRob1994b}%
  \BibitemOpen
  \bibfield  {author} {\bibinfo {author} {\bibfnamefont {T.}~\bibnamefont
  {Prosen}}\ and\ \bibinfo {author} {\bibfnamefont {M.}~\bibnamefont
  {Robnik}},\ }\href@noop {} {\bibfield  {journal} {\bibinfo  {journal} {J.
  Phys. A: Math. Gen.}\ }\textbf {\bibinfo {volume} {27}},\ \bibinfo {pages}
  {8059} (\bibinfo {year} {1994}{\natexlab{b}})}\BibitemShut {NoStop}%
\bibitem [{\citenamefont {Prosen}(1998{\natexlab{a}})}]{Pro1998a}%
  \BibitemOpen
  \bibfield  {author} {\bibinfo {author} {\bibfnamefont {T.}~\bibnamefont
  {Prosen}},\ }\href@noop {} {\bibfield  {journal} {\bibinfo  {journal} {J.
  Phys. A: Math. Gen.}\ }\textbf {\bibinfo {volume} {31}},\ \bibinfo {pages}
  {L345} (\bibinfo {year} {1998}{\natexlab{a}})}\BibitemShut {NoStop}%
\bibitem [{\citenamefont {Prosen}(1998{\natexlab{b}})}]{Pro1998}%
  \BibitemOpen
  \bibfield  {author} {\bibinfo {author} {\bibfnamefont {T.}~\bibnamefont
  {Prosen}},\ }\href@noop {} {\bibfield  {journal} {\bibinfo  {journal} {J.
  Phys. A: Math. Gen.}\ }\textbf {\bibinfo {volume} {31}},\ \bibinfo {pages}
  {7023} (\bibinfo {year} {1998}{\natexlab{b}})}\BibitemShut {NoStop}%
\bibitem [{\citenamefont {Grossmann}\ and\ \citenamefont
  {Robnik}(2007{\natexlab{a}})}]{GroRob1}%
  \BibitemOpen
  \bibfield  {author} {\bibinfo {author} {\bibfnamefont {S.}~\bibnamefont
  {Grossmann}}\ and\ \bibinfo {author} {\bibfnamefont {M.}~\bibnamefont
  {Robnik}},\ }\href@noop {} {\bibfield  {journal} {\bibinfo  {journal} {J.
  Phys. A: Math. Theor.}\ }\textbf {\bibinfo {volume} {40}},\ \bibinfo {pages}
  {409} (\bibinfo {year} {2007}{\natexlab{a}})}\BibitemShut {NoStop}%
\bibitem [{\citenamefont {Grossmann}\ and\ \citenamefont
  {Robnik}(2007{\natexlab{b}})}]{GroRob2}%
  \BibitemOpen
  \bibfield  {author} {\bibinfo {author} {\bibfnamefont {S.}~\bibnamefont
  {Grossmann}}\ and\ \bibinfo {author} {\bibfnamefont {M.}~\bibnamefont
  {Robnik}},\ }\href@noop {} {\bibfield  {journal} {\bibinfo  {journal} {Z.
  Naturforschung A}\ }\textbf {\bibinfo {volume} {62}},\ \bibinfo {pages} {471}
  (\bibinfo {year} {2007}{\natexlab{b}})}\BibitemShut {NoStop}%
\bibitem [{\citenamefont {Batisti\'c}\ and\ \citenamefont
  {Robnik}(2010)}]{BatRob2010}%
  \BibitemOpen
  \bibfield  {author} {\bibinfo {author} {\bibfnamefont {B.}~\bibnamefont
  {Batisti\'c}}\ and\ \bibinfo {author} {\bibfnamefont {M.}~\bibnamefont
  {Robnik}},\ }\href@noop {} {\bibfield  {journal} {\bibinfo  {journal} {J.
  Phys. A: Math. Gen.}\ }\textbf {\bibinfo {volume} {43}},\ \bibinfo {pages}
  {215101} (\bibinfo {year} {2010})}\BibitemShut {NoStop}%
\bibitem [{\citenamefont {Batisti\'c}\ \emph {et~al.}(2013)\citenamefont
  {Batisti\'c}, \citenamefont {Manos},\ and\ \citenamefont {Robnik}}]{BMR2013}%
  \BibitemOpen
  \bibfield  {author} {\bibinfo {author} {\bibfnamefont {B.}~\bibnamefont
  {Batisti\'c}}, \bibinfo {author} {\bibfnamefont {T.}~\bibnamefont {Manos}}, \
  and\ \bibinfo {author} {\bibfnamefont {M.}~\bibnamefont {Robnik}},\
  }\href@noop {} {\bibfield  {journal} {\bibinfo  {journal} {Europhys. Lett.}\
  }\textbf {\bibinfo {volume} {102}},\ \bibinfo {pages} {50008} (\bibinfo
  {year} {2013})}\BibitemShut {NoStop}%
\bibitem [{\citenamefont {Batisti\'c}\ and\ \citenamefont
  {Robnik}(2013{\natexlab{a}})}]{BatRob2013}%
  \BibitemOpen
  \bibfield  {author} {\bibinfo {author} {\bibfnamefont {B.}~\bibnamefont
  {Batisti\'c}}\ and\ \bibinfo {author} {\bibfnamefont {M.}~\bibnamefont
  {Robnik}},\ }\href@noop {} {\bibfield  {journal} {\bibinfo  {journal} {J.
  Phys. A: Math. Theor.}\ }\textbf {\bibinfo {volume} {46}},\ \bibinfo {pages}
  {315102} (\bibinfo {year} {2013}{\natexlab{a}})}\BibitemShut {NoStop}%
\bibitem [{\citenamefont {Batisti\'c}\ and\ \citenamefont
  {Robnik}(2013{\natexlab{b}})}]{BatRob2013A}%
  \BibitemOpen
  \bibfield  {author} {\bibinfo {author} {\bibfnamefont {B.}~\bibnamefont
  {Batisti\'c}}\ and\ \bibinfo {author} {\bibfnamefont {M.}~\bibnamefont
  {Robnik}},\ }\href@noop {} {\bibfield  {journal} {\bibinfo  {journal} {Phys.
  Rev. E}\ }\textbf {\bibinfo {volume} {88}},\ \bibinfo {pages} {052913}
  (\bibinfo {year} {2013}{\natexlab{b}})}\BibitemShut {NoStop}%
\bibitem [{\citenamefont {Vidmar}\ \emph {et~al.}(2007)\citenamefont {Vidmar},
  \citenamefont {St\"ockmann}, \citenamefont {Robnik}, \citenamefont {Kuhl},
  \citenamefont {H\"ohmann},\ and\ \citenamefont {Grossmann}}]{VSRKHG}%
  \BibitemOpen
  \bibfield  {author} {\bibinfo {author} {\bibfnamefont {G.}~\bibnamefont
  {Vidmar}}, \bibinfo {author} {\bibfnamefont {H.~J.}\ \bibnamefont
  {St\"ockmann}}, \bibinfo {author} {\bibfnamefont {M.}~\bibnamefont {Robnik}},
  \bibinfo {author} {\bibfnamefont {U.}~\bibnamefont {Kuhl}}, \bibinfo {author}
  {\bibfnamefont {R.}~\bibnamefont {H\"ohmann}}, \ and\ \bibinfo {author}
  {\bibfnamefont {S.}~\bibnamefont {Grossmann}},\ }\href@noop {} {\bibfield
  {journal} {\bibinfo  {journal} {J. Phys. A: Math. Theor.}\ }\textbf {\bibinfo
  {volume} {40}},\ \bibinfo {pages} {13883} (\bibinfo {year}
  {2007})}\BibitemShut {NoStop}%
\bibitem [{\citenamefont {Prosen}(2000)}]{Pro2000}%
  \BibitemOpen
  \bibfield  {author} {\bibinfo {author} {\bibfnamefont {T.}~\bibnamefont
  {Prosen}},\ }\href@noop {} {}\bibinfo {howpublished} {\textit{in Proceedings
  of the International School of Physics ``Enrico Fermi'', Course CXLIII}}
  (\bibinfo {year} {2000}),\ \bibinfo {note} {edited by G. Casati and I.
  Guarneri and U. Smilyanski (Amsterdam: IOS Press, 2000) p.~473}\BibitemShut
  {NoStop}%
\bibitem [{\citenamefont {Brody}(1973)}]{Bro1973}%
  \BibitemOpen
  \bibfield  {author} {\bibinfo {author} {\bibfnamefont {T.~A.}\ \bibnamefont
  {Brody}},\ }\href@noop {} {\bibfield  {journal} {\bibinfo  {journal} {Lett.
  Nuovo Cimento}\ }\textbf {\bibinfo {volume} {7}},\ \bibinfo {pages} {482}
  (\bibinfo {year} {1973})}\BibitemShut {NoStop}%
\bibitem [{\citenamefont {Brody}\ \emph {et~al.}(1981)\citenamefont {Brody},
  \citenamefont {Flores}, \citenamefont {French}, \citenamefont {Mello},
  \citenamefont {Pandey},\ and\ \citenamefont {Wong}}]{Bro1981}%
  \BibitemOpen
  \bibfield  {author} {\bibinfo {author} {\bibfnamefont {T.~A.}\ \bibnamefont
  {Brody}}, \bibinfo {author} {\bibfnamefont {J.}~\bibnamefont {Flores}},
  \bibinfo {author} {\bibfnamefont {J.~B.}\ \bibnamefont {French}}, \bibinfo
  {author} {\bibfnamefont {P.~A.}\ \bibnamefont {Mello}}, \bibinfo {author}
  {\bibfnamefont {A.}~\bibnamefont {Pandey}}, \ and\ \bibinfo {author}
  {\bibfnamefont {S.~S.~M.}\ \bibnamefont {Wong}},\ }\href@noop {} {\bibfield
  {journal} {\bibinfo  {journal} {Rev. Mod. Phys.}\ }\textbf {\bibinfo {volume}
  {53}},\ \bibinfo {pages} {385} (\bibinfo {year} {1981})}\BibitemShut
  {NoStop}%
\bibitem [{\citenamefont {Izrailev}(1988)}]{Izr1988}%
  \BibitemOpen
  \bibfield  {author} {\bibinfo {author} {\bibfnamefont {F.~M.}\ \bibnamefont
  {Izrailev}},\ }\href@noop {} {\bibfield  {journal} {\bibinfo  {journal}
  {Phys. Lett. A}\ }\textbf {\bibinfo {volume} {134}},\ \bibinfo {pages} {13}
  (\bibinfo {year} {1988})}\BibitemShut {NoStop}%
\bibitem [{\citenamefont {Izrailev}(1990)}]{Izr1990}%
  \BibitemOpen
  \bibfield  {author} {\bibinfo {author} {\bibfnamefont {F.~M.}\ \bibnamefont
  {Izrailev}},\ }\href@noop {} {\bibfield  {journal} {\bibinfo  {journal}
  {Phys. Rep.}\ }\textbf {\bibinfo {volume} {196}},\ \bibinfo {pages} {299}
  (\bibinfo {year} {1990})}\BibitemShut {NoStop}%
\bibitem [{\citenamefont {Manos}\ and\ \citenamefont {Robnik}(2013)}]{MR2013}%
  \BibitemOpen
  \bibfield  {author} {\bibinfo {author} {\bibfnamefont {T.}~\bibnamefont
  {Manos}}\ and\ \bibinfo {author} {\bibfnamefont {M.}~\bibnamefont {Robnik}},\
  }\href@noop {} {\bibfield  {journal} {\bibinfo  {journal} {Phys. Rev. E}\
  }\textbf {\bibinfo {volume} {87}},\ \bibinfo {pages} {062905} (\bibinfo
  {year} {2013})}\BibitemShut {NoStop}%
\bibitem [{\citenamefont {Manos}\ and\ \citenamefont {Robnik}(2014)}]{MR2014}%
  \BibitemOpen
  \bibfield  {author} {\bibinfo {author} {\bibfnamefont {T.}~\bibnamefont
  {Manos}}\ and\ \bibinfo {author} {\bibfnamefont {M.}~\bibnamefont {Robnik}},\
  }\href@noop {} {\bibfield  {journal} {\bibinfo  {journal} {Phys. Rev. E}\
  }\textbf {\bibinfo {volume} {89}},\ \bibinfo {pages} {022905} (\bibinfo
  {year} {2014})}\BibitemShut {NoStop}%
\bibitem [{\citenamefont {Manos}\ and\ \citenamefont {Robnik}(2015)}]{MR2015}%
  \BibitemOpen
  \bibfield  {author} {\bibinfo {author} {\bibfnamefont {T.}~\bibnamefont
  {Manos}}\ and\ \bibinfo {author} {\bibfnamefont {M.}~\bibnamefont {Robnik}},\
  }\href@noop {} {\bibfield  {journal} {\bibinfo  {journal} {Phys. Rev. E}\
  }\textbf {\bibinfo {volume} {91}},\ \bibinfo {pages} {042904} (\bibinfo
  {year} {2015})}\BibitemShut {NoStop}%
\bibitem [{\citenamefont {Izrailev}(1986)}]{Izr1986}%
  \BibitemOpen
  \bibfield  {author} {\bibinfo {author} {\bibfnamefont {F.~M.}\ \bibnamefont
  {Izrailev}},\ }\href@noop {} {\bibfield  {journal} {\bibinfo  {journal}
  {Phys. Rev. Lett.}\ }\textbf {\bibinfo {volume} {56}},\ \bibinfo {pages}
  {541} (\bibinfo {year} {1986})}\BibitemShut {NoStop}%
\bibitem [{\citenamefont {Izrailev}(1987)}]{Izr1987}%
  \BibitemOpen
  \bibfield  {author} {\bibinfo {author} {\bibfnamefont {F.~M.}\ \bibnamefont
  {Izrailev}},\ }\href@noop {} {\bibfield  {journal} {\bibinfo  {journal}
  {Phys. Lett. A}\ }\textbf {\bibinfo {volume} {125}},\ \bibinfo {pages} {250}
  (\bibinfo {year} {1987})}\BibitemShut {NoStop}%
\bibitem [{\citenamefont {Izrailev}(1989)}]{Izr1989}%
  \BibitemOpen
  \bibfield  {author} {\bibinfo {author} {\bibfnamefont {F.~M.}\ \bibnamefont
  {Izrailev}},\ }\href@noop {} {\bibfield  {journal} {\bibinfo  {journal} {J.
  Phys. A: Math. Gen.}\ }\textbf {\bibinfo {volume} {22}},\ \bibinfo {pages}
  {865} (\bibinfo {year} {1989})}\BibitemShut {NoStop}%
\bibitem [{\citenamefont {Shepelyansky}(1986)}]{She1986}%
  \BibitemOpen
  \bibfield  {author} {\bibinfo {author} {\bibfnamefont {D.~L.}\ \bibnamefont
  {Shepelyansky}},\ }\href@noop {} {\bibfield  {journal} {\bibinfo  {journal}
  {Phys. Rev. Lett.}\ }\textbf {\bibinfo {volume} {56}},\ \bibinfo {pages}
  {677} (\bibinfo {year} {1986})}\BibitemShut {NoStop}%
\bibitem [{\citenamefont {Taylor}(1969)}]{T69}%
  \BibitemOpen
  \bibfield  {author} {\bibinfo {author} {\bibfnamefont {J.~B.}\ \bibnamefont
  {Taylor}},\ }\href@noop {} {} (\bibinfo {year} {1969}),\ \bibinfo {note}
  {culham Laboratory Progress Report, CLM-PR-12 (1969)}\BibitemShut {NoStop}%
\bibitem [{\citenamefont {Froeschl\'e}(1970)}]{F72}%
  \BibitemOpen
  \bibfield  {author} {\bibinfo {author} {\bibfnamefont {C.}~\bibnamefont
  {Froeschl\'e}},\ }\href@noop {} {\bibfield  {journal} {\bibinfo  {journal}
  {Astron. Astrophys.}\ }\textbf {\bibinfo {volume} {9}},\ \bibinfo {pages}
  {15} (\bibinfo {year} {1970})}\BibitemShut {NoStop}%
\bibitem [{\citenamefont {Chirikov}(1979)}]{C79}%
  \BibitemOpen
  \bibfield  {author} {\bibinfo {author} {\bibfnamefont {B.}~\bibnamefont
  {Chirikov}},\ }\href@noop {} {\bibfield  {journal} {\bibinfo  {journal}
  {Phys. Rep.}\ }\textbf {\bibinfo {volume} {52}},\ \bibinfo {pages} {263}
  (\bibinfo {year} {1979})}\BibitemShut {NoStop}%
\bibitem [{\citenamefont {Lichtenberg}\ and\ \citenamefont
  {Lieberman}(1992)}]{LL1992}%
  \BibitemOpen
  \bibfield  {author} {\bibinfo {author} {\bibfnamefont {A.~J.}\ \bibnamefont
  {Lichtenberg}}\ and\ \bibinfo {author} {\bibfnamefont {M.~A.}\ \bibnamefont
  {Lieberman}},\ }\href@noop {} {\emph {\bibinfo {title} {Regular and Chaotic
  Dynamics}}}\ (\bibinfo  {publisher} {New York: Springer Verlag},\ \bibinfo
  {year} {1992})\BibitemShut {NoStop}%
\bibitem [{\citenamefont {Skokos}\ \emph {et~al.}(2007)\citenamefont {Skokos},
  \citenamefont {Bountis},\ and\ \citenamefont {Antonopoulos}}]{SBA:2007}%
  \BibitemOpen
  \bibfield  {author} {\bibinfo {author} {\bibfnamefont {C.}~\bibnamefont
  {Skokos}}, \bibinfo {author} {\bibfnamefont {T.}~\bibnamefont {Bountis}}, \
  and\ \bibinfo {author} {\bibfnamefont {C.}~\bibnamefont {Antonopoulos}},\
  }\href@noop {} {\bibfield  {journal} {\bibinfo  {journal} {Physica D}\
  }\textbf {\bibinfo {volume} {231}},\ \bibinfo {pages} {30} (\bibinfo {year}
  {2007})}\BibitemShut {NoStop}%
\bibitem [{\citenamefont {Skokos}\ and\ \citenamefont {Manos}(2015)}]{SM2014}%
  \BibitemOpen
  \bibfield  {author} {\bibinfo {author} {\bibfnamefont {C.}~\bibnamefont
  {Skokos}}\ and\ \bibinfo {author} {\bibfnamefont {T.}~\bibnamefont {Manos}},\
  }\href@noop {} {\bibfield  {journal} {\bibinfo  {journal} {(e-preprint:
  arXiv:1412.7401)}\ } (\bibinfo {year} {2015})}\BibitemShut {NoStop}%
\bibitem [{\citenamefont {Zeldovich}(1966)}]{Zel1966}%
  \BibitemOpen
  \bibfield  {author} {\bibinfo {author} {\bibfnamefont {Y.~B.}\ \bibnamefont
  {Zeldovich}},\ }\href@noop {} {\bibfield  {journal} {\bibinfo  {journal}
  {Eksp. Teor. Fiz.}\ }\textbf {\bibinfo {volume} {51}},\ \bibinfo {pages}
  {1942} (\bibinfo {year} {1966})}\BibitemShut {NoStop}%
\bibitem [{\citenamefont {Izrailev}\ and\ \citenamefont
  {Shepelyansky}(1979{\natexlab{a}})}]{IS1979a}%
  \BibitemOpen
  \bibfield  {author} {\bibinfo {author} {\bibfnamefont {F.~M.}\ \bibnamefont
  {Izrailev}}\ and\ \bibinfo {author} {\bibfnamefont {D.~L.}\ \bibnamefont
  {Shepelyansky}},\ }\href@noop {} {\bibfield  {journal} {\bibinfo  {journal}
  {Dokl. Akad. Nauk SSSR}\ }\textbf {\bibinfo {volume} {249}},\ \bibinfo
  {pages} {1103} (\bibinfo {year} {1979}{\natexlab{a}})}\BibitemShut {NoStop}%
\bibitem [{\citenamefont {Izrailev}\ and\ \citenamefont
  {Shepelyansky}(1979{\natexlab{b}})}]{IS1979b}%
  \BibitemOpen
  \bibfield  {author} {\bibinfo {author} {\bibfnamefont {F.~M.}\ \bibnamefont
  {Izrailev}}\ and\ \bibinfo {author} {\bibfnamefont {D.~L.}\ \bibnamefont
  {Shepelyansky}},\ }\href@noop {} {\bibfield  {journal} {\bibinfo  {journal}
  {Sov. Phys. Dokl.}\ }\textbf {\bibinfo {volume} {24}},\ \bibinfo {pages}
  {996} (\bibinfo {year} {1979}{\natexlab{b}})}\BibitemShut {NoStop}%
\bibitem [{\citenamefont {Izrailev}\ and\ \citenamefont
  {Shepelyansky}(1980{\natexlab{a}})}]{IS1980a}%
  \BibitemOpen
  \bibfield  {author} {\bibinfo {author} {\bibfnamefont {F.~M.}\ \bibnamefont
  {Izrailev}}\ and\ \bibinfo {author} {\bibfnamefont {D.~L.}\ \bibnamefont
  {Shepelyansky}},\ }\href@noop {} {\bibfield  {journal} {\bibinfo  {journal}
  {Teor. Mat. Fiz.}\ }\textbf {\bibinfo {volume} {43}},\ \bibinfo {pages} {417}
  (\bibinfo {year} {1980}{\natexlab{a}})}\BibitemShut {NoStop}%
\bibitem [{\citenamefont {Izrailev}\ and\ \citenamefont
  {Shepelyansky}(1980{\natexlab{b}})}]{IS1980b}%
  \BibitemOpen
  \bibfield  {author} {\bibinfo {author} {\bibfnamefont {F.~M.}\ \bibnamefont
  {Izrailev}}\ and\ \bibinfo {author} {\bibfnamefont {D.~L.}\ \bibnamefont
  {Shepelyansky}},\ }\href@noop {} {\bibfield  {journal} {\bibinfo  {journal}
  {Theor. Math. Phys.}\ }\textbf {\bibinfo {volume} {43}},\ \bibinfo {pages}
  {553} (\bibinfo {year} {1980}{\natexlab{b}})}\BibitemShut {NoStop}%
\bibitem [{\citenamefont {Casati}\ and\ \citenamefont
  {Guarneri}(1984)}]{CG1984}%
  \BibitemOpen
  \bibfield  {author} {\bibinfo {author} {\bibfnamefont {G.}~\bibnamefont
  {Casati}}\ and\ \bibinfo {author} {\bibfnamefont {I.}~\bibnamefont
  {Guarneri}},\ }\href@noop {} {\bibfield  {journal} {\bibinfo  {journal}
  {Commun. Math. Phys.}\ }\textbf {\bibinfo {volume} {95}},\ \bibinfo {pages}
  {121} (\bibinfo {year} {1984})}\BibitemShut {NoStop}%
\bibitem [{\citenamefont {Chirikov}\ \emph {et~al.}(1981)\citenamefont
  {Chirikov}, \citenamefont {Izrailev},\ and\ \citenamefont
  {Shepelyansky}}]{CIS1981}%
  \BibitemOpen
  \bibfield  {author} {\bibinfo {author} {\bibfnamefont {B.~V.}\ \bibnamefont
  {Chirikov}}, \bibinfo {author} {\bibfnamefont {F.~M.}\ \bibnamefont
  {Izrailev}}, \ and\ \bibinfo {author} {\bibfnamefont {D.~L.}\ \bibnamefont
  {Shepelyansky}},\ }\href@noop {} {\bibfield  {journal} {\bibinfo  {journal}
  {Sov. Sci. Revv. C 2}\ }\textbf {\bibinfo {volume} {2}},\ \bibinfo {pages}
  {209} (\bibinfo {year} {1981})}\BibitemShut {NoStop}%
\bibitem [{\citenamefont {Casati}\ \emph {et~al.}(1990)\citenamefont {Casati},
  \citenamefont {Guarneri}, \citenamefont {Izrailev},\ and\ \citenamefont
  {Scharf}}]{CGIS1990}%
  \BibitemOpen
  \bibfield  {author} {\bibinfo {author} {\bibfnamefont {G.}~\bibnamefont
  {Casati}}, \bibinfo {author} {\bibfnamefont {I.}~\bibnamefont {Guarneri}},
  \bibinfo {author} {\bibfnamefont {F.~M.}\ \bibnamefont {Izrailev}}, \ and\
  \bibinfo {author} {\bibfnamefont {R.}~\bibnamefont {Scharf}},\ }\href@noop {}
  {\bibfield  {journal} {\bibinfo  {journal} {Phys. Rev. Lett.}\ }\textbf
  {\bibinfo {volume} {64}},\ \bibinfo {pages} {5} (\bibinfo {year}
  {1990})}\BibitemShut {NoStop}%
\bibitem [{\citenamefont {Kottos}\ \emph {et~al.}(1996)\citenamefont {Kottos},
  \citenamefont {Politi}, \citenamefont {Izrailev},\ and\ \citenamefont
  {Ruffo}}]{Kottos1996}%
  \BibitemOpen
  \bibfield  {author} {\bibinfo {author} {\bibfnamefont {T.}~\bibnamefont
  {Kottos}}, \bibinfo {author} {\bibfnamefont {A.}~\bibnamefont {Politi}},
  \bibinfo {author} {\bibfnamefont {F.}~\bibnamefont {Izrailev}}, \ and\
  \bibinfo {author} {\bibfnamefont {S.}~\bibnamefont {Ruffo}},\ }\href@noop {}
  {\bibfield  {journal} {\bibinfo  {journal} {Phys. Rev. E}\ }\textbf {\bibinfo
  {volume} {53}},\ \bibinfo {pages} {R5553} (\bibinfo {year}
  {1996})}\BibitemShut {NoStop}%
\bibitem [{\citenamefont {Fishman}\ \emph {et~al.}(1982)\citenamefont
  {Fishman}, \citenamefont {Grempel},\ and\ \citenamefont {Prange}}]{FGP1982}%
  \BibitemOpen
  \bibfield  {author} {\bibinfo {author} {\bibfnamefont {S.}~\bibnamefont
  {Fishman}}, \bibinfo {author} {\bibfnamefont {D.}~\bibnamefont {Grempel}}, \
  and\ \bibinfo {author} {\bibfnamefont {R.}~\bibnamefont {Prange}},\
  }\href@noop {} {\bibfield  {journal} {\bibinfo  {journal} {Phys. Rev. Lett.}\
  }\textbf {\bibinfo {volume} {49}},\ \bibinfo {pages} {509} (\bibinfo {year}
  {1982})}\BibitemShut {NoStop}%
\bibitem [{\citenamefont {Prange}\ \emph {et~al.}(1984)\citenamefont {Prange},
  \citenamefont {Grempel},\ and\ \citenamefont {Fishman}}]{PGF1984}%
  \BibitemOpen
  \bibfield  {author} {\bibinfo {author} {\bibfnamefont {R.}~\bibnamefont
  {Prange}}, \bibinfo {author} {\bibfnamefont {D.}~\bibnamefont {Grempel}}, \
  and\ \bibinfo {author} {\bibfnamefont {S.}~\bibnamefont {Fishman}},\
  }\href@noop {} {\emph {\bibinfo {title} {Como Conference on Quantum Chaos, G.
  Casati, ed.}}}\ (\bibinfo  {publisher} {Plenum, New York},\ \bibinfo {year}
  {1984})\BibitemShut {NoStop}%
\bibitem [{\citenamefont {Fujisaka}(1983)}]{Fuj1983}%
  \BibitemOpen
  \bibfield  {author} {\bibinfo {author} {\bibfnamefont {H.}~\bibnamefont
  {Fujisaka}},\ }\href@noop {} {\bibfield  {journal} {\bibinfo  {journal}
  {Prog. Theor. Phys.}\ }\textbf {\bibinfo {volume} {70}},\ \bibinfo {pages}
  {1264} (\bibinfo {year} {1983})}\BibitemShut {NoStop}%
\bibitem [{\citenamefont {Ott}(1993)}]{Ott1993}%
  \BibitemOpen
  \bibfield  {author} {\bibinfo {author} {\bibfnamefont {E.}~\bibnamefont
  {Ott}},\ }\href@noop {} {\emph {\bibinfo {title} {Chaos in Dynamical
  Systems}}}\ (\bibinfo  {publisher} {Cambridge University Press},\ \bibinfo
  {year} {1993})\BibitemShut {NoStop}%
\bibitem [{\citenamefont {Kottos}\ \emph {et~al.}(1999)\citenamefont {Kottos},
  \citenamefont {Izrailev},\ and\ \citenamefont {Politi}}]{Kottos1999}%
  \BibitemOpen
  \bibfield  {author} {\bibinfo {author} {\bibfnamefont {T.}~\bibnamefont
  {Kottos}}, \bibinfo {author} {\bibfnamefont {F.}~\bibnamefont {Izrailev}}, \
  and\ \bibinfo {author} {\bibfnamefont {A.}~\bibnamefont {Politi}},\
  }\href@noop {} {\bibfield  {journal} {\bibinfo  {journal} {Physica D}\
  }\textbf {\bibinfo {volume} {131}},\ \bibinfo {pages} {155} (\bibinfo {year}
  {1999})}\BibitemShut {NoStop}%
\end{thebibliography}%

\end{document}